%% file: manuscript.tex
\documentclass[aip,jcp,11pt,onecolumn,tightlines,superscriptaddress,preprintnumbers]{revtex4}  
\usepackage{amsfonts}
\usepackage{amsmath}
\usepackage{amssymb}
\usepackage{listings}
\usepackage{version}
\usepackage{color}
\usepackage{hyperref}
\usepackage{appendix}
\usepackage{pgfplots}
\usepackage{listings}
\usepackage[capitalize]{cleveref}
\usepackage{graphicx}%
\includeversion{New_connection}
\excludeversion{Old_connection}

\renewcommand{\L}{\mathcal{L}}

\usepgfplotslibrary{units}

\begin{document}
\preprint{To appear in Journal of Chemical Physics -- DOI: 10.1063/5.0066008 -- LA-UR-21-27590}

\title{Bringing discrete-time Langevin splitting methods into agreement with thermodynamics}

\author{Joshua Finkelstein}
\email{jdf@lanl.gov}
\affiliation{Theoretical Division, Los Alamos National Laboratory, Los Alamos, NM 87545, USA}

\author{Chungho Cheng}
\email{argcheng@ucdavis.edu}
\affiliation{Department of Mechanical \& Aerospace Engineering,
  University of California, Davis, CA 95616, USA}
  
\author{Giacomo Fiorin}
\email{giacomo.fiorin@temple.edu}
\altaffiliation[Current address: ]{National Institute of Neurological Disorders and Stroke, Bethesda, MD 20892, USA}
\affiliation{Institute for Computational Molecular Science, Temple University, Philadelphia, PA 19122, USA}

\author{Benjamin Seibold}
\email{seibold@temple.edu}
\affiliation{Department of Mathematics, Temple University, Philadelphia, PA 19122, USA}
  
\author{Niels Gr{\o}nbech-Jensen}
\email{ngjensen@math.ucdavis.edu}
\thanks{Corresponding author}
\affiliation{Department of Mechanical \& Aerospace Engineering, University of California, Davis, CA 95616, USA}
\affiliation{Department of Mathematics, University of California, Davis, CA 95616, USA}

\begin{abstract}

\begin{center}
{\Large Abstract}
\end{center}

    In light of the recently published complete set of statistically correct Gr{\o}nbech-Jensen (GJ) methods for discrete-time thermodynamics, we revise a differential operator splitting method for the Langevin equation in order to comply with the basic GJ thermodynamic sampling features, namely the Boltzmann distribution and Einstein diffusion, in linear systems. This revision, which is based on the introduction of time scaling along with flexibility of a discrete-time velocity attenuation parameter, provides a direct link between the ABO splitting formalism and the GJ methods. This link brings about the conclusion that any GJ method has at least weak second order accuracy in the applied time step. It further helps identify a novel half-step velocity, which simultaneously produces both correct kinetic statistics and correct transport measures for any of the statistically sound GJ methods. Explicit algorithmic expressions are given for the integration of the new half-step velocity into the GJ set of methods. Numerical simulations, including quantum-based molecular dynamics (QMD) using the QMD suite LATTE, highlight the discussed properties of the algorithms as well as exhibit the direct application of robust, time step independent stochastic integrators to quantum-based molecular dynamics.
\noindent  
\end{abstract}
\maketitle

\section{Introduction}
\label{sec:intro}
    Langevin dynamics \cite{Langevin} is an oft-used approach for investigating thermodynamic systems, including Molecular Dynamics (MD) \cite{AllenTildesley,Frenkel,Rapaport,Hoover_book,Leach}, in the canonical ensemble of, say, $N$ particles, each with coordinate $r_i$, associated velocity $v_i$, mass $m_i$, subject to a force $f_i=-\nabla_{r_i}U(\{r_j\})$, where $U(\{r_j\})$ is the potential energy of the system configuration $\{r_j\}$, and a positive friction constant $\alpha_i$, which couples the coordinates to a common thermal bath with thermodynamic temperature $T$. The statistical sampling of the fluctuation-dissipation balance for particles in three dimensions can be modeled by the stochastic Langevin equation \cite{Langevin,Langevin_Eq} 
    \begin{eqnarray}
        m_i\,\ddot{r}_i & = & f_i-\alpha_i\,\dot{r}_i+\beta_i\label{eq:Langevin}
    \end{eqnarray}
    \noindent{}or as the first order system
    \begin{subequations}
    \begin{eqnarray}
    dr_i & = & v_i\,dt\\
    m_i\,dv_i & = & f_i\,dt-\alpha_iv_i\,dt+\beta_i\,dt\, ,
    \end{eqnarray}
    \label{eq:Langevin_2}
    \end{subequations}
    \noindent{}where $i=1,2,\cdots,N$, $r_i=(r_{i,x},r_{i,y},r_{i,z})^T$, $v_i=(v_{i,x},v_{i,y},v_{i,z})^T$ and where dissipation is balanced by the fluctuation term $\beta_i$ if \cite{Parisi,Langevin_Eq}
    \begin{subequations}
    \begin{eqnarray}
    \langle\beta_i(t)\rangle & = & 0 \\
    \langle\beta_{i,\mu}(t)\beta_{j,\nu}(s)\rangle & = & 2\alpha_i\,k_BT\,\delta(t-s)\,\delta_{ij}\,\delta_{\mu\nu} \;,
    \end{eqnarray}\label{eq:FD}
    \end{subequations}
    $k_B$ being Boltzmann's constant, and $\mu,\nu=x,y,z$.

    Computational statistical mechanics has made extensive use of the Langevin equation and many discrete-time algorithms have been proposed for simulations, especially ones based on the deterministic St{\o}rmer-Verlet method \cite{Verlet,Stormer_1921,Newton}, which is accurate to second order in the time step size. Thus, while many algorithms, both stochastic \cite{Martyna_92,Watanabe,SS,vgb_1982,Gunsteren,BBK,Pastor_88,Bussi_2007,Bussi_07} and deterministic \cite{Berendsen_84,Nose,Hoover,Hoover2}, have previously been developed and used, they have typically included intrinsic time step errors in the sampling statistics, even for linear systems. This feature is not surprising given the systematic discrete-time errors, which typically accumulate differently through the deterministic and stochastic terms of a stochastic differential equation. Therefore, simulations of the Langevin equation have generally necessitated very small time steps (relative to the stability limit) if statistical results are to be obtained with reasonable accuracy. 

    The past decade has seen significant progress in the development and deployment of reliable discrete-time stochastic algorithms for accurate, efficient, and robust simulations of the Langevin equation \cite{LM,GJF1,Sivak,2GJ,GJ}. The new paradigm of obtaining {\it correct} statistics from methods with inherent time step errors was introduced in Refs.~\cite{LM,GJF1}, even when using large time steps. The first method to demonstrate this for a harmonic potential was the BAOAB method of the Leimkuhler-Matthews (LM) family \cite{LM}, which exactly reproduces Boltzmann statistics for a harmonic oscillator at any time step within the stability range. This method, however, exhibits systematic numerical errors in transport properties, such as Einstein diffusion and drift in a flat potential. The GJF method \cite{GJF1} was developed to produce correct, time-step-independent statistics for both the harmonic and flat potentials, thus also providing correct transport in discrete time. Numerous direct simulations of both simple nonlinear and complex molecular dynamics systems have since confirmed these attractive features (see, e.g., Refs.~\cite{GJF2,GJF4,Li2017,2GJ,GJ,Josh,JDF2}) and two variants of this method \cite{GJF1,2GJ} are included in the public distribution of the molecular dynamics simulation suite LAMMPS \cite{Plimpton,LAMMPS-Manual}.

    The two methods, BAOAB and GJF, were developed in very different ways. The BAOAB method is derived from a Strang splitting \cite{Strang} of differential operators obtained from the Langevin equation, as has also been previously pursued in, e.g., Ref.~\cite{rc_2003}. The GJF method was originally derived in a more ad-hoc manner from direct integration of the Langevin equation with subsequent approximations that preserve temporal symmetry. An interesting further development to the LM BAOAB method was later shown in Ref.~\cite{Sivak} to correct the diffusive properties through a particular rescaling of the time step in BAOAB, thus creating an algorithm with the same overall statistical properties as GJF. This revision to the BAOAB method was named VRORV~\cite{Sivak}. More recently, in Ref.~\cite{GJ}, it was shown that the GJF and the VRORV methods in fact both belong to an exclusive class of similar methods, the Gr{\o}nbech-Jensen (GJ) methods, which all exactly reproduce Boltzmann statistics for harmonic potentials {\it and} provide exact drift and Einstein diffusion for a flat potential. This set was derived through an analytical  brute-force parameter matching approach, which achieved the desired time-step-independent Boltzmann distribution, drift, and Einstein diffusion. Additionally, it was shown that stochastic Verlet-type methods {\it cannot} provide correct statistics for both configurational and kinetic sampling from variables evaluated at the same discrete times. However, kinetic statistics {\it can} be correctly sampled from the half-steps between two successive time steps of a statistically correct configurational coordinate \cite{GJ}, and such half-step velocities were defined as a generalization of the previously identified, statistically correct 2GJ half-step velocity \cite{2GJ}. Most recently, Ref.~\cite{JDF2} has explicitly shown that a stochastic Verlet-type method {\it cannot} produce correct time-step-independent configurational statistics if more than one stochastic number is applied to each degree of freedom per time step. This crucial result underlines that the set of GJ methods is, indeed, a complete and exclusive set of methods that have the above mentioned desired statistical properties.

    The GJF and VRORV methods, corresponding to the GJ-I and GJ-II methods of the GJ set listed in Ref.~\cite{GJ} when applying the same half-step velocity estimator to both, have similar statistical properties to one another. However, the derivations of these methods differ in a fundamental way since VRORV originated from a corrective time scaling of the BAOAB method which, in turn, originated from the operating splitting technique. We seek in this paper to identify corrective measures to the splitting approach that can similarly lead to all methods in the GJ family. The reason for this interest is two-fold. First, such revisions would provide insight to the relationships between numerical Langevin methods and methodologies. Second, with a direct relationship between the methods and how they are derived, one may be able to take advantage of the analysis framework developed for one approach in the analysis of the methods arising from the other.

    Building on the ABO splitting technique advanced by Ref.~\cite{LM}, and combined with the insight of the specific corrective deterministic time scaling highlighted in Ref.~\cite{Sivak} for a specific case, we are in
    Sec.~\ref{sec:general-splitting} able to consistently generalize the ABO splitting terms in order to reproduce the GJ set of methods. In Sec.~\ref{sec:equivalence} we explain how our generalized ABO formalism leads to the realization that all GJ methods are equivalent to the BAOAB method, up to a rescaling of friction. Given that BAOAB is known to be weak second order, i.e., it produces steady-state distributions that deviate at most by terms second order in the time step for any non-linear system, this suggests that the same is true for all GJ methods. An explicit calculation in support of this idea using the already-developed analytical machinery for the BAOAB splitting technique is presented in the Appendix.
    
    From this combined analysis, we further identify a new GJ method, which is the result of equal time scaling throughout all three A, B, and O splitting operations. Finally, in Sec.~\ref{subsec:half-step} we revisit the issue of the definition of a statistically correct half-step velocity, and identify a general expression that can be applied to any of the GJ methods with correct Maxwell-Boltzmann as well as transport statistics. We provide explicit algorithmic integration of this new half-step velocity into the GJ set. 
    The results are validated in Sec.~\ref{sec:Num_Sim} through i) numerical simulations of simple non-linear systems including direct tests of the weak second order accuracy, and ii) quantum-based molecular dynamics (QMD) to show the usefulness of both the GJ methods for QMD and the new statistically robust half-step velocity in quantifying the simulation's adherence to equilibrium Boltzmann and Maxwell-Boltzmann statistics. We note that the GJ methods have previously been extensively tested on nonlinear and classical MD systems; see, e.g., Refs.~\cite{GJF2,GJF4,2GJ,Josh,GJ,JDF2}.

\section{Discrete-time algorithms}
\label{sec:background}

    In this section we review the basic desirable statistical requirements that a discrete-time stochastic algorithm should deliver for its configurational (i.e., spatial) coordinate in a linear system. These are (i) the spatial autocorrelation function for a harmonic potential, (ii) the diffusion constant for a flat potential, and (iii) the steady drift for a constant external force. Because the only connectivity between the particles is through the interaction force $f_i$, and the same subscript is on every variable and parameter in the equation, we can without ambiguity omit the subscript $i$ in Eqs.~(\ref{eq:Langevin})--(\ref{eq:FD}). Thus, in the following we will include the subscript $i$ only if it is necessary for clarity of presentation. Likewise, we only include the Cartesian indices $\mu,\nu=x,y,z$ when necessary.

    We seek the discrete-time approximations $r^n$ to the continuous time variables $r$ at times $t_n=t_0+n\,\Delta{t}$, for initial time $t_0$ and the subsequent $n$ steps of discrete time $\Delta{t}$. For the harmonic oscillator, $f=-\kappa r$ ($\kappa \ge 0$), which has natural frequency $\Omega_0=\sqrt{\kappa/m}$, and the flat potential given by $\kappa=0$, the desired results are that
    \begin{subequations}
    \begin{align}
    \langle r_\mu^nr_\mu^n\rangle \; & = &\frac{k_BT}{\kappa} & & {\rm for} \; \kappa>0 \label{eq:Boltz} \\
    \lim_{n\,\Delta{t}\rightarrow\infty}\frac{\left\langle(r_\mu^n-r_\mu^0)^2\right\rangle}{2\,n\,\Delta{t}} \; & = &\frac{k_BT}{\alpha} & \; = \; D_E & {\rm for} \; \kappa=0\label{eq:Einst}\\
    \frac{\langle r^{n+1}-r^n\rangle}{\Delta{t}} & = & \frac{f}{\alpha}& & {\rm for} \; f^n=f={\rm const.}, \kappa=0 \label{eq:Drift}
    \end{align}\label{eq:Stats}
    \end{subequations}
    with $\mu=x,y,z$, for any discrete time step $\Delta{t}$. These expressions represent the correct Boltzmann distribution in Eq.~(\ref{eq:Boltz}), the correct Einstein diffusion in Eq.~(\ref{eq:Einst}), and the correct drift velocity in Eq.~(\ref{eq:Drift}). It is important to emphasize that these three conditions, which are investigated for linear systems, are mutually independent. For example, the BBK method \cite{BBK} satisfies Eqs.~((\ref{eq:Einst}) and (\ref{eq:Drift}), while not satisfying Eq.~(\ref{eq:Boltz}), the BAOAB method \cite{LM} satisfies Eq.~(\ref{eq:Boltz}), while not satisfying Eqs.~(\ref{eq:Einst}) and (\ref{eq:Drift}), and the GJF and GJ methods \cite{GJF1,GJ} satisfy all three conditions. For nonlinear problems, a method that satisfies Eq.~(\ref{eq:Boltz}), may still deviate from exact Boltzmann statistics, however it is self-evident that a method that is correct in the linear limit is preferable to one that is not. As described in Refs.~\cite{GJ} these conditions lead to an exclusive and complete set of stochastic Verlet-type methods that can be identified without explicit representation of an associated discrete-time velocity. Discrete-time velocities can subsequently be constructed to yield the correct kinetic response, as has also been demonstrated in Refs.~\cite{2GJ,GJ}. The following two subsections give a brief review of the two types of methods that we investigate; first those in the set of GJ methods, which includes GJF, and then the BAOAB method.

\subsection{The set of thermodynamically correct GJ methods}
\label{subsec:GJ}

    The GJ set of stochastic Verlet-type methods for approximating Eq.~(\ref{eq:Langevin_2}) with time step $\Delta{t}$ such that Eq.~(\ref{eq:Stats}) is satisfied, can be written, e.g., in the velocity-Verlet form \cite{GJ}
    \begin{subequations}
    \begin{eqnarray}
        r^{n+1} & = & r^n+\sqrt{c_1c_3}\,\Delta{t}\,v^n+\frac{c_3\Delta{t}^2}{2m}\,f^n+\frac{c_3\Delta{t}}{2m}\,\beta^{n+1} \label{eq:GJ_r}\\
        v^{n+1} & = & c_2\,v^n+\sqrt{\frac{c_3}{c_1}}\frac{\Delta{t}}{2m}(c_2\,f^n+f^{n+1})+\frac{\sqrt{c_1c_3}}{m}\,\beta^{n+1}\label{eq:GJ_v}\, ,
    \end{eqnarray}\label{eq:GJ}
    \end{subequations}
    where the discrete-time on-site velocity, $v^n$, has been introduced, and the single noise contribution for each degree of freedom during a time step is
    \begin{eqnarray}
        \beta^{n+1} & = & \int_{t_n}^{t_{n+1}}\beta\,dt
    \end{eqnarray}
    with $\beta$ being the continuous-time white noise given by Eq.~(\ref{eq:FD}). The cumulative discrete-time noise is thus
    \begin{eqnarray}
        \beta_{i,\mu}^{n} & = & \sqrt{2\alpha\,k_BT\,\Delta{t}}\,\sigma_{i,\mu}^{n}\label{eq:beta}
    \end{eqnarray}
    on particle $i$, in direction $\mu$, over a time step $\Delta{t}$ for $\sigma_i^n$ being a unit Gaussian random variable, satisfying
    \begin{subequations}
    \begin{eqnarray}
        \langle\sigma_i^n\rangle & = & 0 \\
        \langle\sigma_{i,\mu}^n\sigma_{j,\nu}^\ell\rangle & = & \delta_{n\ell}\,\delta_{ij}\,\delta_{\mu\nu}\, .
    \end{eqnarray}\label{eq:noise_n}
    \end{subequations}
    The functional parameters $c_1$ and $c_3$ are
    \begin{subequations}
    \begin{eqnarray}
        2\,c_1 & = & 1+c_2 \label{eq:c1}\\
    \gamma\Delta{t}\,c_3 & = & 1-c_2 \label{eq:c3}
    \end{eqnarray}\label{eq:c13}
    \end{subequations}
    with $m\gamma=\alpha$, and $c_2=c_2(\gamma\Delta{t})$ is the single time step velocity attenuation parameter (function) that defines the specific GJ method. Equation~(\ref{eq:GJ_v}) requires $|c_2|\le 1$, and Eq.~(\ref{eq:c3}) requires that $c_2\sim (1-\gamma\Delta{t})$ as $\gamma\Delta{t}\rightarrow0$. It is assumed that $c_2$ is a decaying function for which $c_2\rightarrow1$ for $\gamma\Delta{t}\rightarrow0$. Three key examples of GJ methods were given in Ref.~\cite{GJ}
    \begin{subequations}
    \begin{eqnarray}
    {\rm GJ\!-\!I \; (GJF)}: & c_2 \; = & \frac{1-\frac{1}{2}\gamma\Delta{t}}{1+\frac{1}{2}\gamma\Delta{t}} \; = \; a \; \Rightarrow \; c_1=c_3 \label{eq:c2_gjf}\\
    {\rm GJ\!-\!II \; (VRORV)}: & c_2 \; =  & \exp(-\gamma\Delta{t})  \label{eq:c2_vrorv} \\
    {\rm GJ\!-\!III}: & c_2 \; =  & 1-\gamma\Delta{t}  \; \Rightarrow \; c_3=1 \qquad ({\rm when } \; \; \gamma \Delta t < 2)\label{eq:c2_gj3}\, .
    \end{eqnarray}
    \end{subequations}

    It was found that for convex potentials the on-site velocity $v^n$, defined from $\langle v^nr^n\rangle=0$, given in Eq.~(\ref{eq:GJ_v}), results in calculated kinetic energies smaller in magnitude than the correct values \cite{GJF1}. However the corresponding half-step velocity, denoted by $u^{n+\frac{1}{2}}$ and following from the anti-symmetry requirement $\langle u^{n+\frac{1}{2}}r^n\rangle+\langle u^{n-\frac{1}{2}}r^n\rangle=0$, is defined to be
    \begin{eqnarray}
    u^{n+\frac{1}{2}} & = & \frac{r^{n+1}-r^n}{\sqrt{c_3}\,\Delta{t}} \label{eq:GJ-u}\, ,
    \end{eqnarray}
    and was found in Refs.~\cite{2GJ,GJ} to yield the correct kinetic result
    \begin{eqnarray}
    \langle u_\mu^{n+\frac{1}{2}} u_\mu^{n+\frac{1}{2}}\rangle & = & \frac{k_BT}{m} \label{eq:Maxw}
    \end{eqnarray}
    for linear systems with $f=-\kappa r$ ($\kappa\ge0$), regardless of time step $\Delta{t}$, within the zero temperature stability limit:
    \begin{eqnarray}
    \Omega_0^2\Delta{t}^2 & < & 4\frac{c_1}{c_3}\, . \label{eq:GJ_stability}
    \end{eqnarray}
    As noted in Ref.~\cite{GJ}, the drift properties of the half-step velocity $u$ in Eq.~(\ref{eq:GJ-u}) are correct only for the GJ-III method, since $c_3=1$ in that case.

    A main goal of this paper is to relate the GJ methods, which are {\it the only stochastic Verlet-type integrators that can satisfy the essential configurational sampling results given in Eq.~(\ref{eq:Stats})}, to the differential operator splitting technique while using the same parameter-search technique to enforce the statistical objectives of the resulting methods.

\subsection{ABO splitting and the BAOAB method}
\label{subsec:ABO}

    Intuitively, we can view Langevin splitting methods as first starting from \cref{eq:Langevin_2}, breaking the vector field up into its simpler pieces, solving each piece exactly, and then recombining these solutions to yield an approximate method. Given an initial phase space distribution $\rho_0$, this distribution is evolved over time such that the distribution at time $t$ is given by the expression $\rho_t = e^{t \mathcal{L}^*_{\rm LD}} \rho_0$, where $\mathcal{L}^*_{\rm LD}$ is the Fokker-Planck operator for Langevin dynamics (LD). This is analogous to the explicit writing of the Langevin equation given in Eq.~(\ref{eq:Langevin_2}). The evolution of the system can be partitioned into three characteristic components \cite{LM}: inertial (A), interactive (B), and thermodynamic (O), as outlined below in Eq.~(\ref{eq:Split_1}) for both the Fokker-Planck operator in (\ref{eq:FP_split}) and the corresponding Langevin terms in (\ref{eq:Langevin_split}) \cite{LM_book}:

    \begin{subequations}
    \begin{align}
        \mathcal{L}^*_{\rm LD} &= &  \overbrace{-\sum_i v_i\cdot\nabla_{r_i}}^{\mathcal{L}^*_A} & \quad + &  \overbrace{-\sum_i \frac{1}{m_i}f_i\cdot\nabla_{v_i}}^{\mathcal{L}^*_B} 
       & \quad + &\overbrace{\sum_i \gamma_i ( \nabla_{v_i}v_i\cdot +  \frac{k_BT }{m_i}\,\Delta_{v_i})}^{\mathcal{L}^*_O}&  \label{eq:FP_split} \\
       d\left(\begin{array}{c}r_i\\v_i\end{array}\right) & = & \underbrace{\left(\begin{array}{c}v_i\,dt\\0\end{array}\right)}_{A}\; \; \; &\quad +&\underbrace{\left(\begin{array}{c}0\\ \frac{1}{m_i}f_i\,dt\end{array}\right)}_{B}\; \; \; & \quad + &\underbrace{\left(\begin{array}{c}0\\ -\gamma_i v_i\,dt+\frac{1}{m_i}\beta_i\,dt\end{array}\right)}_{O}\; \; \; & \label{eq:Langevin_split}\, .
    \end{align}\label{eq:Split_1}
    \end{subequations}

    Here, $m_i\gamma_i=\alpha_i$. Since $e^{t \mathcal{L}^*_{\rm LD}}$ is not easily calculable, the partitioning, or splitting, method attempts to isolate more easily computable pieces in order to combine the results and approximate $e^{t \mathcal{L}^*_{\rm LD}}$. For instance, if we write $\mathcal{L}^*_{\rm LD} = \mathcal{L}_{\rm A}^* + \mathcal{L}_{\rm OB}^*$ with $\mathcal{L}^*_{\rm OB}=\mathcal{L}^*_{\rm O}+\mathcal{L}^*_{\rm B}$, we can make the approximation $e^{t \mathcal{L}^*_{\rm LD}} \approx e^{t \mathcal{L}_{\rm A}^*} e^{t \mathcal{L}_{\rm OB}^*}$, which was the splitting order used in Ref.~\cite{rc_2003}. This approximation is exact only when $[ \mathcal{L}_{\rm A}^*, \mathcal{L}_{\rm OB}^*]=0$, which can be viewed as a consequence of the Baker-Campbell-Hausdorff (BCH) expansion (see, e.g., Ref.~\cite{LM_book} Sec.~3.1 and~7.7). We can equivalently view this splitting as having started with \cref{eq:Langevin_split}, breaking the vector field into A and OB pieces with each piece solved exactly, and then composing the resulting solution operators in the appropriate order. Another significant splitting approach was used in Ref.~\cite{Bussi_2007} to combine the A and B operators into a single operation AB, such that ${\mathcal{L}^*_{\rm LD}} = \mathcal{L}_{\rm AB}^* + \mathcal{L}_{\rm O}^*$ with $\mathcal{L}^*_{\rm AB}=\mathcal{L}^*_{\rm A}+\mathcal{L}^*_{\rm B}$. The work of Ref.~\cite{LM} that we connect to in this paper kept the three given operations as separately evaluated contributions ${\mathcal{L}^*_{\rm LD}} = \mathcal{L}_{\rm A}^* + \mathcal{L}_{\rm B}^*+ \mathcal{L}_{\rm O}^*$. For a non-zero time interval $\Delta{t}$, the solution operators for the A and B operations are evaluated analytically and become standard Euler approximations. Similarly the O operation (the thermodynamic Ornstein-Uhlenbeck step) can also be evaluated analytically. The resulting effects of the Fokker-Planck operators in discrete-time for the particle coordinates can then be written:
    \begin{subequations}
    \begin{eqnarray}
        e^{\Delta{t}\mathcal{L}^*_A} \; : & \begin{pmatrix} r \\ v \end{pmatrix} & {\overset{A}{\longrightarrow}} \; \begin{pmatrix} r+\Delta{t}\,v \\ v\end{pmatrix} \label{eq:Split_A}\\
        e^{\Delta{t}\mathcal{L}^*_B} \; : & \begin{pmatrix} r \\ v \end{pmatrix} & {\overset{B}{\longrightarrow}} \; \begin{pmatrix} r \\ v+\frac{\Delta{t}}{m}f\end{pmatrix} \label{eq:Split_B}\\
        e^{\Delta{t}\mathcal{L}^*_O} \; : & \begin{pmatrix} r \\ v \end{pmatrix} & {\overset{O}{\longrightarrow}} \; \begin{pmatrix} r \\ c_2\,v+\sqrt{c_1c_3}\frac{1}{m}\beta\end{pmatrix} \; = \; \begin{pmatrix} r \\ c_2\,v+\sqrt{(1-c_2^2)\frac{k_BT}{m}}\,\sigma\end{pmatrix} \label{eq:Split_O}\, ,
    \end{eqnarray}\label{eq:Split_ABO}
    \end{subequations}
    where $\sigma$ is an {\it i.i.d.}, $\delta$-correlated standard normal distribution, and where 
    \begin{eqnarray}
        c_2=\exp(-\gamma\Delta{t})\label{eq:c2_baoab}
    \end{eqnarray}
    coincides with Eq.~(\ref{eq:c2_vrorv}). We stress that the exponential operators on the left-hand side in Eq.~(\ref{eq:Split_ABO}) act on distributions, and not on phase space points. The notation in Eq.~(\ref{eq:Split_ABO}) is a convenience where the operator \emph{samples} the evolved distribution when starting from a $\delta$-distribution peaked at $(r, v)$. Thus the composition of operators in Eq.~(\ref{eq:Split_ABO}) corresponds to the sampling of a stochastic path in phase space from a known initial condition. The associated parameters $c_1$ and $c_3$ are given by Eq.~(\ref{eq:c13}). The B step assumes that $f=f(r)$, which implies that the A step is understood to be followed by an updated evaluation of $f$ before another B step is conducted. The operations in Eq.~(\ref{eq:Split_ABO}) imply that
    \begin{subequations}
    \begin{eqnarray}
        e^{\frac{1}{2}\Delta{t}\mathcal{L}^*_A}\,e^{\frac{1}{2}\Delta{t}\mathcal{L}^*_A}\, & = & e^{\Delta{t}\mathcal{L}^*_A}\,\label{eq:AA}\\
        e^{\frac{1}{2}\Delta{t}\mathcal{L}^*_B}\,e^{\frac{1}{2}\Delta{t}\mathcal{L}^*_B}\, & = & e^{\Delta{t}\mathcal{L}^*_B}\, \label{eq:BB}\\
        e^{\frac{1}{2}\Delta{t}\mathcal{L}^*_O}\,e^{\frac{1}{2}\Delta{t}\mathcal{L}^*_O}\, & = & e^{\Delta{t}\mathcal{L}^*_O}\, \label{eq:OO}\, ,
    \end{eqnarray}
    \end{subequations}
    where the exponential form of Eq.~(\ref{eq:c2_baoab}) is necessary for Eq.~(\ref{eq:OO}). Even if Eq.~(\ref{eq:Split_O}) indicates that the two consecutive applications of the thermodynamic operation on the left hand side of Eq.~(\ref{eq:OO}) produces two different random numbers, $\sigma_+$ and $\sigma_-$, these contributions combine to produce the single random variable $\sigma$ for the completed time step, $t_n\rightarrow t_{n+1}$:
    \begin{subequations}
    \begin{eqnarray}
        \sigma^{n+1} & = & \frac{\sqrt{c_2}\sigma_+^n+\sigma_-^{n+1}}{\sqrt{1+c_2}}\label{eq:sigma_onsite_split}
    \end{eqnarray}
    such that
    \begin{eqnarray}
        \beta^{n+1} & = & \sqrt{2\alpha\,k_BT\,\Delta{t}}\,\sigma^{n+1}\label{eq:beta_onsite_split}\, ,
    \end{eqnarray}
    \end{subequations}
    where $\sigma_\pm^n$ are unit Gaussian random variables. Leimkuhler and Matthews \cite{LM} thus symmetrically applied the above half-time step operations in order to obtain overall temporal symmetry and minimize distributional error among the other ABO methods,
    \begin{eqnarray}
        {\rm BAOAB} & \longleftrightarrow & e^{\frac{1}{2}\Delta{t}\mathcal{L}^*_B}\,e^{\frac{1}{2}\Delta{t}\mathcal{L}^*_A}\,\underbrace{e^{\frac{1}{2}\Delta{t}\mathcal{L}^*_O}\,e^{\frac{1}{2}\Delta{t}\mathcal{L}^*_O}}_{e^{\Delta{t}\mathcal{L}^*_O}}\,e^{\frac{1}{2}\Delta{t}\mathcal{L}^*_A}\,e^{\frac{1}{2}\Delta{t}\mathcal{L}^*_B}\, ,
    \end{eqnarray}
    leading to the so-called BAOAB method given explicitly as
    \begin{subequations}
    \begin{eqnarray}
        r^{n+1} & = & r^n+c_1\,\Delta{t}\,v^n+\frac{c_1\Delta{t}^2}{2m}\,f^n+\frac{\sqrt{c_1c_3}\Delta{t}}{2m}\,\beta^{n+1} \label{eq:BAOAB_r}\\
        v^{n+1} & = & c_2\,v^n+\frac{\Delta{t}}{2m}(c_2\,f^n+f^{n+1})+\frac{\sqrt{c_1c_3}}{m}\,\beta^{n+1}\label{eq:BAOAB_v}\, ,
    \end{eqnarray}\label{eq:BAOAB}
    \end{subequations}
    where $c_2$ is given by Eq.~(\ref{eq:c2_baoab}), and $c_1$ and $c_3$ are given in Sec.~\ref{subsec:GJ}. For linear systems with $f=-\kappa r$, this method satisfies Eq.~(\ref{eq:Boltz}) when $\kappa>0$, but for $\kappa=0$ the Einstein diffusion and the drift velocity yield
    \begin{eqnarray}
        \lim_{n\,\Delta{t}\rightarrow\infty}\frac{\left\langle(r_\mu^n-r_\mu^0)^2\right\rangle}{2\,n\,\Delta{t}} \; & = & \frac{k_BT}{\alpha} \, \sqrt{\frac{c_1}{c_3}} \; \; \;  {\rm for} \; f=0 \label{eq:Einst_BAOAB}\\
        \frac{\langle r^{n+1}-r^n\rangle}{\Delta{t}} & = & \frac{c_1}{c_3}\,\frac{f}{\alpha}  \; \; \;  {\rm for} \; f={\rm const}\neq0 \label{eq:Drift_BAOAB}\, ,
    \end{eqnarray}
  {\it which are not the desired values given in Eqs.~(\ref{eq:Einst}) and (\ref{eq:Drift}). Thus, while the BAOAB method correctly samples the Boltzmann distribution, it does not correctly describe transport}. Combining the BAOAB method Eq.~(\ref{eq:BAOAB}) with the tailored half-step velocity
    \begin{eqnarray}
        u^{n+\frac{1}{2}} & = & \frac{r^{n+1}-r^n}{\sqrt{c_1}\,\Delta{t}} \label{eq:BAOAB-u}
    \end{eqnarray}
    satisfies the equipartition of velocities in Eq.~(\ref{eq:Maxw}), at least for linear systems with $f=-\kappa r$ ($\kappa\ge0$).

    Comparing Eq.~(\ref{eq:GJ}) with Eq.~(\ref{eq:BAOAB}), Eqs.~(\ref{eq:Einst}) and (\ref{eq:Drift}) with Eqs.~(\ref{eq:Einst_BAOAB}) and (\ref{eq:Drift_BAOAB}), and Eq.~(\ref{eq:GJ-u}) with Eq.~(\ref{eq:BAOAB-u}), it is clear that BAOAB would be equivalent to one of the methods in the GJ set if $c_1=c_3$. This implies $c_2=a$, as is seen in Eq.~(\ref{eq:c2_gjf}) instead of the exponential form given by Eq.~(\ref{eq:c2_baoab}) that arises from the splitting technique reviewed in this section. Thus, if one were to {\it define} the operation in Eq.~(\ref{eq:Split_O}) with $c_2=a$ for the time step $\Delta{t}$, then the GJ-I (GJF-2GJ) method would arise from the splitting technique instead of BAOAB. A similar observation was made previously in \cite{Li2017} where it was shown that by choosing $\gamma$ and $\Delta t$ such that $a = \exp(-\gamma \Delta t)$, BAOAB and GJF could be made to generate the exact same discrete-time trajectories.

    Another approach for mitigating the diffusion and transport problems in BAOAB was proposed in Ref.~\cite{Sivak} by introducing a temporal scaling
    \begin{eqnarray} \label{vrorv time scaling}
        \Delta{t} & \rightarrow & \sqrt{\frac{2}{\gamma\Delta{t}}\tanh\frac{\gamma\Delta{t}}{2}}\,\Delta{t}
    \end{eqnarray}
    in the A and B operations shown in Eqs.~(\ref{eq:Split_A}) and (\ref{eq:Split_B}), while maintaining the exponential form for $c_2$ in the O operation of Eq.~(\ref{eq:Split_O}). This readily leads to the VRORV method, which coincides in its configurational statistics with the GJ-II method given by Eq.~(\ref{eq:c2_vrorv}) in Sec.~\ref{subsec:GJ}.

\section{Generalized Splitting}\label{sec:general-splitting}

    The previous section suggests that the ABO splitting method needs at least one of two revisions in its operations before it can conform to the criteria defining the GJ methods, and before all the thermodynamic quantities Eqs.~(\ref{eq:Stats}) and (\ref{eq:Maxw}) can be satisfied. One possibility is to revise the one-time-step velocity attenuation parameter $c_2$; the other is to introduce scaling of the time step in operations A and B (similar to Ref.~\cite{Sivak}). Here, we combine the above-mentioned observations to revise the splitting operations A, B, and O in Eq.~(\ref{eq:Split_ABO}), in order to arrive at the entire GJ set of methods through two key generalizations.

    The first generalization is to allow for a non-exponential, one-time-step velocity attenuation parameter $c_2$ in the O operation. The thermodynamic O operation is only temporally additive as shown in Eqs.~(\ref{eq:Split_O}) and (\ref{eq:OO}) if $c_2$ has the exponential form of Eq.~(\ref{eq:c2_baoab}). One way to address this is to {\it define} the thermodynamic additivity $e^{\frac{1}{2}\Delta{t}\mathcal{L}^*_O}e^{\frac{1}{2}\Delta{t}\mathcal{L}^*_O}=e^{\Delta{t}\mathcal{L}^*_O}$ from an assumed relationship $(c_2(\frac{1}{2}\gamma\Delta{t}))^2=c_2(\gamma\Delta{t})$.
    The configurational sampling is, however, not limited by the additivity of the O operation, since Ref.~\cite{JDF2} has shown that no stochastic Verlet-type integrator can produce correct Boltzmann statistics with more than one independent stochastic number contributing to the configurational equation per time step. If O were not at the pivotal (i.e., central) position, but instead used as two symmetrically positioned $\frac{1}{2}\Delta{t}$ operations, then the configurational variable $r^{n+1}$ would be subject to two separate stochastic contributions, $\sigma_+^n$ and $\sigma_-^{n+1}$. Thus, the thermodynamic operation O must be placed at the central position in the sequence of splitting steps with a full $\Delta{t}$ advancement of time. We therefore adopt Eq.~(\ref{eq:Split_O}), using any allowed functional parameter $c_2=c_2(-\gamma\Delta{t})$.

    The second generalization is to introduce temporal scaling parameters, $d_A$, $d_B$, $d_O$, throughout each of the A, B, and O operations in Eq.~(\ref{eq:Langevin_split}). We write the generalized operations $\widetilde{\rm A}$, $\widetilde{\rm B}$, and $\widetilde{\rm O}\widetilde{\rm O}$ 
    \begin{subequations}
    \begin{eqnarray}
    e^{\frac{1}{2}\Delta{t}\widetilde{\mathcal{L}}^*_A} \; : & \begin{pmatrix} r \\ v \end{pmatrix} & {\overset{\widetilde{A}}{\longrightarrow}} \; \begin{pmatrix} r+d_A\frac{\Delta{t}}{2}\,v \\ v\end{pmatrix} \label{eq:Split_A_rev}\\
    e^{\frac{1}{2}\Delta{t}\widetilde{\mathcal{L}^*_B}} \; : & \begin{pmatrix} r \\ v \end{pmatrix} & {\overset{\widetilde{B}}{\longrightarrow}} \; \begin{pmatrix} r \\ v+d_B\frac{\Delta{t}}{2m}f\end{pmatrix} \label{eq:Split_B_rev}\\
    e^{\Delta{t}\widetilde{\mathcal{L}}^*_O} \; : & \begin{pmatrix} r \\ v \end{pmatrix} & {\overset{\widetilde{O}\widetilde{O}}{\longrightarrow}} \; \begin{pmatrix} r \\ c_2\,v+\sqrt{(1-c_2^2)\frac{k_BT}{m}}\,\sigma\end{pmatrix} \label{eq:Split_O_rev}\, ,
    \end{eqnarray}\label{eq:Split_ABO_rev}
    \end{subequations}
    where $c_2=c_2(d_O\gamma\Delta{t})$ must satisfy the conditions outlined for a one-time-step velocity attenuation parameter in Sec.~\ref{subsec:GJ}, $c_1$ and $c_3$ are given by Eqs.~(\ref{eq:c13}), and $\sigma$ is a unit Gaussian as described in Sec.~\ref{subsec:ABO}. The reason for the double $\widetilde{\rm O}$, instead of a single $\widetilde{\rm O}$, in Eq.~(\ref{eq:Split_O_rev}) will be made clear in the next section. Arranging these generalized operators in Eq.~(\ref{eq:Split_ABO_rev}) into a BAOAB-like ordering, or equivalently a velocity-Verlet ordering with a pivotal random term, leads to a generalized BAOAB-type method, which when written in exponential operator form is expressed as
    \begin{align} \label{generic_operator_splitting_form}
    \begin{split}
        e^{\frac{\Delta t}{2}\widetilde{\mathcal{L}}_B^*} e^{\frac{\Delta t}{2}\widetilde{\mathcal{L}}_A^*} & e^{ \Delta t \widetilde{\mathcal{L}}_O^*} e^{\frac{\Delta t}{2}\widetilde{\mathcal{L}}_A^*} e^{\frac{\Delta t}{2}\widetilde{\mathcal{L}}_B^*} \\
        &=
        e^{d_{B}\frac{\Delta t}{2}\mathcal{L}_B^*} e^{d_{A}\frac{\Delta t}{2}\mathcal{L}_A^*} e^{d_{O} \Delta t \mathcal{L}_O^*} e^{d_{A}\frac{\Delta t}{2}\mathcal{L}_A^*} e^{d_{B}\frac{\Delta t}{2}\mathcal{L}_B^*} \;.
    \end{split}
    \end{align} 
    Equivalently, at the $n$-th step, the right-hand side of (\ref{generic_operator_splitting_form}) can be expressed using the more direct form,
    \begin{align}
    \begin{split}
        v^{n+1/4} &=  v^n+d_B\frac{\Delta{t}}{2m}f^n\\
        r^{n+1/2} &= r^n+d_A\frac{\Delta{t}}{2}\,v^{n+1/4}\\
        v^{n+3/4} &= c_2\,v^{n+1/4}+\sqrt{(1-c_2^2){k_BT}{m}^{-1}} \; \sigma^{n+1} \\
        r^{n+1} &=  r^{n+1/2}+d_A\frac{\Delta{t}}{2}\,v^{n+3/4}\\
        v^{n+1} &= v^{n+3/4}+d_B\frac{\Delta{t}}{2m}f^{n+1}\;,
    \end{split}\label{eq:gen_split}
    \end{align}
    where $v^{n+\frac{1}{4}}$ and $v^{n+\frac{3}{4}}$ are intermediate variables that do not correspond to the fractional time step indicated by the superscript. Further, rewriting (\ref{eq:gen_split}) leads to the explicit update rule
    \begin{subequations}
    \begin{eqnarray}
        r^{n+1} & = & r^n+d_Ac_1\,\Delta{t}\,v^n+d_Ad_B\frac{c_1\Delta{t}^2}{2m}\,f^n+d_A\frac{\sqrt{c_1c_3}\Delta{t}}{2m}\,\beta^{n+1} \label{eq:BAOAB_r_rev}\\
        v^{n+1} & = & c_2\,v^n+d_B\frac{\Delta{t}}{2m}(c_2\,f^n+f^{n+1})+\frac{\sqrt{c_1c_3}}{m}\,\beta^{n+1}\, .\label{eq:BAOAB_v_rev}
    \end{eqnarray}\label{eq:BAOAB_rev}
    \end{subequations}
    Comparing Eq.~(\ref{eq:BAOAB_rev}) with Eq.~(\ref{eq:GJ}) shows that the two temporal scaling parameters $d_A$ and $d_B$ must be chosen equal and identical to:
    \begin{eqnarray}
    d_{A} \; = \; d_B & = & d_{AB} \; = \; \sqrt{\frac{c_3}{c_1}} = \; \sqrt{\frac{2}{\gamma \Delta t}\frac{1-c_2}{1+c_2}}\label{eq:dAdB}\;.
    \end{eqnarray}
    For $c_2$ specifically given by Eq.~(\ref{eq:c2_baoab}), we are led directly to the time scaling given by Eq.~(\ref{vrorv time scaling}) and hence to the VRORV method described in Ref.~\cite{Sivak}.
    
    The temporal scaling parameter $d_O$ in the thermodynamic operation $\widetilde{\rm O}$ is not explicitly given by the enforced conditions. This scaling is simply embedded in the choice of $c_2$, and it can be chosen freely within the existing GJ family of methods for as long as $c_2$ satisfies its basic conditions listed in Sec.~\ref{subsec:GJ} (see also Ref.~\cite{GJ}). Since those conditions imply that the derived $c_1$ and $c_3$ both have a limit of unity for $\gamma\Delta{t}\rightarrow0$, it also follows that $d_{AB} \to 1$ for $\gamma\Delta{t}\rightarrow0$. This then implies that for any allowed velocity attenuation parameter, $c_2(\gamma\Delta{t})$, the function $c_2(d_{AB}\gamma\Delta{t})$ is a valid one-time-step velocity attenuation parameter in the GJ set, since the functional parameter can be viewed as merely scaling the friction parameter $\gamma$. Thus, a rather straightforward way of extending the time scaling $d_{AB}$ to the set of ABO splitting methods is simply to choose $d_O=d_{AB}$:
    \begin{subequations}
    \begin{eqnarray}
    c_2 & = & \exp\left(-\sqrt{\frac{c_3}{c_1}}\,\gamma\Delta{t}\right) \label{eq:c2_GJ-Josh} \\
    \gamma\Delta{t} & = & \frac{1+c_2}{1-c_2}\,\frac{(\ln{c_2})^2}{2}\label{eq:dt_GJ-Josh}
    \end{eqnarray}\label{eq:GJ-Josh}
    \end{subequations}
    for $0<c_2\le1$. We denote the method defined by $c_2$ in Eq.~(\ref{eq:GJ-Josh}) the GJ-UTS method for universal time scaling. For consistency with the enumeration of six other GJ methods in Ref.~\cite{GJ}, we also label this method as GJ-VII. In comparison with earlier methods of the GJ set, we notice that for the GJ-I method, where $c_2=a$ is given by Eq.~(\ref{eq:c2_gjf}), $c_2(d_{AB}\gamma\Delta{t})=c_2(\gamma\Delta{t})=a$ is invariant to a $d_O=d_{AB}$ temporal scaling in the thermodynamic operation, Eq.~(\ref{eq:Split_O_rev}).

\subsection{Half-step velocities} \label{subsec:half-step}

    As mentioned above, an essential feature of the pivotal location of the thermodynamic step O is that the two independent noise terms from the two sequential half-time steps OO combine into a single noise contribution $\beta^{n+1}$ for the coordinates $r^n$ and $v^n$ seen in, e.g., Eq.~(\ref{eq:BAOAB_rev}). Inspired by the exponential origins of the splitting methods, we {\it define}, for any allowable functional form of $c_2=c_2(\gamma\Delta{t})$, the generalized thermodynamic half-step operation
    \begin{eqnarray}
        e^{\frac{\Delta{t}}{2}\widetilde{\mathcal{L}}^*_O}\; : & \begin{pmatrix} r \\ v \end{pmatrix} & {\overset{\widetilde{O}}{\longrightarrow}} \; \begin{pmatrix} r \\ \sqrt{c_2}\,v+\sqrt{(1-c_2)\frac{k_BT}{m}}\sigma\end{pmatrix} \label{eq:Split_O_rev_half}\, ,
    \end{eqnarray}
    where $\sigma$ is drawn from a zero mean, unit variance Gaussian distribution. The half-step velocity $u_s^{n+\frac{1}{2}}$, defined by the half-time-step sequence $\widetilde{\rm O}\widetilde{\rm A}\widetilde{\rm B}$ given by Eqs.~(\ref{eq:Split_A_rev}), (\ref{eq:Split_B_rev}), and (\ref{eq:Split_O_rev_half}) on $(r^n,v^n)$, then yields
    \begin{eqnarray}
        u_s^{n+\frac{1}{2}} & = & \sqrt{c_2}\, (v^n+\frac{d_{B}\,\Delta{t}}{2m}f^n)+\sqrt{(1-c_2)\frac{k_BT}{m}}\,\sigma^n_+ \\
        & = & \frac{\sqrt{c_2}}{d_{A}\,c_1}\frac{r^{n+1}-r^n}{\Delta{t}}+\frac{1}{2c_1}\sqrt{(1-c_2^2)\frac{k_BT}{m}}\,\underbrace{\frac{\sigma^n_+-\sqrt{c_2}\sigma^{n+1}_-}{\sqrt{1+c_2}}}_{\sigma^{n+\frac{1}{2}}}\label{eq:us_1}\\
        & = & \frac{\sqrt{c_2}}{d_{A}\,c_1}\frac{r^{n+1}-r^n}{\Delta{t}}+\frac{1}{2m}\sqrt{\frac{c_3}{c_1}}\,\beta^{n+\frac{1}{2}}  \, , \label{eq:us}
    \end{eqnarray}
    where we have used Eq.~(\ref{eq:BAOAB_r_rev}) to arrive at Eq.~(\ref{eq:us_1}), and where the effective half-step fluctuations $\sigma^{n+\frac{1}{2}}$ and $\beta^{n+\frac{1}{2}}$ are defined consistent with the definition of the on-site fluctuations $\sigma^n$ and $\beta^n$ in Sec.~\ref{subsec:ABO}. Choosing $d_A$ and $d_B$ according to Eq.~(\ref{eq:dAdB}) makes this velocity applicable to the GJ methods:
    \begin{eqnarray}
        u_s^{n+\frac{1}{2}} & = & \sqrt{\frac{c_2}{c_1c_3}}\,\frac{r^{n+1}-r^n}{\Delta{t}}+\frac{1}{2m}\sqrt{\frac{c_3}{c_1}}\,\beta^{n+\frac{1}{2}} \label{eq:us_gj}\, .
    \end{eqnarray}

    Since $\langle\sigma^{\ell}\sigma^{n+\frac{1}{2}}\rangle=\langle\beta^\ell\beta^{n+\frac{1}{2}}\rangle=0$, it follows for linear systems that $u_s^{n+\frac{1}{2}}$ satisfies the condition for being a half-step velocity, $\langle u_s^{n+\frac{1}{2}}(r^n+r^{n+1})\rangle=0$ (see Ref.~\cite{GJ}). Additionally, this half-step velocity produces the correct kinetic energy  given by $\langle u_s^{n+\frac{1}{2}}u_s^{n+\frac{1}{2}}\rangle=\frac{k_BT}{m}$. However, it is evident from the prefactor to the first term on the right hand side of Eq.~(\ref{eq:us_gj}) that $u_s^{n+\frac{1}{2}}$ does not correctly express drift and non-stochastic transport for $\alpha>0$. In fact, using this velocity measure, drift for a constant force $f$ is given by
    \begin{eqnarray}
        \langle u_s^{n+\frac{1}{2}}\rangle & = & \frac{\sqrt{c_2}}{d_{A}\,c_1} \, \frac{f}{\alpha} \\
        & = & \sqrt{\frac{c_2}{c_1c_3}}\,\, \frac{f}{\alpha} \; \; , \; \; \; {\rm for} \; d_A=\sqrt{\frac{c_3}{c_1}}\, ,
    \end{eqnarray}
    where the first factor on the right hand side is the discrepancy from the correct result.

    The general treatment above shows that the two independent noise contributions from the two applications of the successive thermodynamic half-step operations $\widetilde{\rm O}$ produce two {\it independent} noise contributions, $\beta^{n+1}$ and $\beta^{n+\frac{1}{2}}$, for on-site and half-step times, respectively. This separation between half-step and on-site noise allows us to propose a novel half-step velocity, which simultaneously satisfies both desirable drift and statistical sampling properties. This velocity applies to any of the GJ methods and can be written
    \begin{align}\label{eq:new_half_step}
        u^{n+\frac{1}{2}}  =  \frac{r^{n+1}-r^n}{\Delta{t}}\pm\frac{1}{m}\sqrt{\frac{1-c_3}{2\frac{\alpha\Delta{t}}{m}}}\,\beta^{n+\frac{1}{2}}\, ,
    \end{align}
    where $\langle\beta^\ell\beta^{n+\frac{1}{2}}\rangle=0$, and where the choice $\pm$ is arbitrary since the noise is Gaussian. For linear systems the velocity has the correct statistical properties regardless of the choices of $d_A$, $d_B$, and $c_2$. Thus, it is a half-step velocity by the anti-symmetric correlation with $r^n$ and $r^{n+1}$; it produces the correct kinetic energy for methods that produce correct configurational statistics; and it gives the correct drift velocity for constant force $f$.
The half-step velocity in Eq.~(\ref{eq:new_half_step}) is derived for the specific form of the GJ methods, and it does therefore not directly apply to BAOAB. Replacing $c_3$ in Eq.~(\ref{eq:new_half_step}) with $c_1$ will make this new velocity applicable to BAOAB.
The BAOAB trajectories combined with the half-step velocity in Eq.~(\ref{eq:us}), for $d_A=d_B=1$, are similar to those that arise from the so-called Langevin ``middle" method \cite{Li2017}. Thus, the GJ methods with the half-step velocity of Eq.~(\ref{eq:new_half_step}) therefore produce better accuracy for both configurational and kinetic measures than what is accomplished by the ``middle scheme".
    
    The new half-step velocity can be incorporated into a leap-frog GJ algorithm as follows
    \begin{subequations}
    \begin{eqnarray}
    u^{n+\frac{1}{2}} & = & c_2\,u^{n-\frac{1}{2}} + \frac{c_3\Delta{t}}{m}\,f^n+\frac{c_3}{2m}(\beta^n+\beta^{n+1})\pm\frac{1}{m}\sqrt{\frac{1-c_3}{2\gamma\Delta{t}}}(\beta^{n+\frac{1}{2}}-c_2\beta^{n-\frac{1}{2}})\\
    r^{n+1} & = & r^n+\Delta{t}\,u^{n+\frac{1}{2}}\mp\frac{\Delta{t}}{m}\sqrt{\frac{1-c_3}{2\gamma\Delta{t}}}\,\beta^{n+\frac{1}{2}}\, ,
    \end{eqnarray}\label{eq:new_GJ_LF}
    \end{subequations}
    and it can also be expressed into the a compact GJ form together with the on-site velocity $v^n$:
   \begin{subequations}
    \begin{eqnarray}
    u^{n+\frac{1}{2}} & = & \sqrt{c_1c_3}\,v^{n} +\frac{c_3\Delta{t}}{2m}\,f^n+\frac{c_3}{2m}\beta^{n+1}\pm\frac{1}{m}\sqrt{\frac{1-c_3}{2\gamma\Delta{t}}}\beta^{n+\frac{1}{2}}\\
    r^{n+1} & = & r^n+\Delta{t}\,u^{n+\frac{1}{2}}\mp\frac{\Delta{t}}{m}\sqrt{\frac{1-c_3}{2\gamma\Delta{t}}}\,\beta^{n+\frac{1}{2}}\\
    v^{n+1} & = & \frac{c_2}{\sqrt{c_1c_3}}\,u^{n+\frac{1}{2}}+\sqrt{\frac{c_3}{c_1}}\frac{\Delta{t}}{2m}\,f^{n+1}+\frac{\sqrt{c_1c_3}}{2m}(2-\frac{c_2}{c_1})\,\beta^{n+1}\pm\frac{1}{m}\frac{c_2}{\sqrt{c_1c_3}}\sqrt{\frac{1-c_3}{2\gamma\Delta{t}}}\,\beta^{n+\frac{1}{2}}\, .
    \end{eqnarray}\label{eq:new_GJ_LF}
    \end{subequations}

    It is worth clarifying that the use of the second, independent noise contribution $\beta^{n+\frac{1}{2}}$ within the time step is exclusively for the half-step velocity, and it is therefore not in conflict with the result of Ref.~\cite{JDF2}, stating that there can only be one stochastic variable per time step for the configurational coordinate $r^n$ to be statistically correct. We notice that this new half step velocity coincides with the previously identified statistically correct half-step velocity for the GJ-III method, where $c_3=1$ \cite{GJ} (see also Eq.~(\ref{eq:GJ-u})).

\section{Method Equivalencies and statistical accuracy} \label{sec:equivalence}

    Equivalencies amongst methods within the GJ set, and also with BAOAB, are very well illuminated by the generalized splitting form outlined above. In exponential operator parlance, the Langevin equation is numerically evolved using a GJ method over a time $\Delta t$ by
    \begin{align} 
        e^{d_{AB}\frac{\Delta t}{2}\mathcal{L}_B^*} e^{d_{AB}\frac{\Delta t}{2}\mathcal{L}_A^*} e^{d_{O} \Delta t \mathcal{L}_O^*} e^{d_{AB}\frac{\Delta t}{2}\mathcal{L}_A^*} e^{d_{AB}\frac{\Delta t}{2}\mathcal{L}_B^*} \;,
    \end{align} 
    which contrasts with the numerical evolution prescribed by BAOAB:
    \begin{align} 
         e^{\frac{\Delta t}{2}\mathcal{L}_B^*} e^{\frac{\Delta t}{2}\mathcal{L}_A^*} e^{ \Delta t \mathcal{L}_O^*} e^{\frac{\Delta t}{2}\mathcal{L}_A^*} e^{\frac{\Delta t}{2}\mathcal{L}_B^*} \;.
    \end{align} 
    The remarks given in the appendix of \cite{Li2017} show that, when $0<\gamma \Delta t<2$, GJF and BAOAB are actually the same method (i.e., they yield identical trajectories and velocities for the same noise) if both methods use the same time step $\Delta t$ but two \emph{different} choices for $\gamma=\alpha/m$. The erroneous diffusion and transport results of the BAOAB method arise because of this rescaling of the damping parameter when comparing to the exact diffusion result $\frac{k_BT}{\alpha}$, where the damping parameter is not rescaled. The GJ methods, including GJF, are constructed to avoid this inconsistency. The viewpoint of friction rescaling between methods can be generalized to include the entire one-parameter family of GJ methods. In this spirit, we may view the splitting method in \cref{generic_operator_splitting_form} to be the BAOAB method with a time step $d_{AB}{\Delta t}$ and a now re-scaled $\gamma$, the scaling factor being the ratio of $d_O$ to $d_{AB}$, i.e.,
    \begin{align} \label{generic_operator_splitting_form_2}
        e^{d_{AB} \frac{ \Delta t}{2}\L^*_B} e^{d_{AB} \frac{\Delta t}{2}\L^*_A} e^{(d_O/d_{AB}) (d_{AB} \Delta t)\L^*_O} e^{ d_{AB} \frac{\Delta t}{2}\L^*_A} e^{ d_{AB} \frac{ \Delta t}{2}\L^*_B} \;,
    \end{align} 
    because
    \begin{align}
        (d_O/d_{AB}) (d_{AB} \Delta t)\L^*_O 
        & = (d_{AB} \Delta t) \sum_i (\tfrac{d_O}{d_{AB}} \gamma) ( \nabla_{v_i}v_i\cdot +  \frac{k_BT }{m_i}\,\Delta_{v_i}) \;.
    \end{align}
    Here, \cref{generic_operator_splitting_form_2} then has the standard splitting interpretation where each portion of the split vector field in \cref{eq:Langevin_split} has the same meaning of time. In this sense, \emph{any} GJ method is mathematically equivalent with BAOAB as well as any other GJ method for some re-scaled $\gamma$ within an appropriate domain of $\gamma \Delta t$. However, unlike BAOAB, the GJ friction rescaling is designed to be consistent with the known expressions for diffusion and drift, Eqs.~(\ref{eq:Einst}) and (\ref{eq:Drift}). We note that the above attribution of the mutual equivalence of the GJ methods to a rescaling of the friction is not necessarily implying that there exists a universally correct way to describe the friction parameter in discrete time; i.e., a universally best choice of GJ method by choosing $c_2$. All GJ methods are designed to correctly give the same basic statistical properties.
    Choosing which method is most appropriate for a given application depends on both the system and on the simulation objectives, which may require particular focus on, e.g., the statistical sampling efficiency or how the simulated frequency components of the system depend on the time step. However it may often be the case that unless warranted for some particular reason, the simplest choice where time is not scaled (i.e., the GJ-I method) may likely be the most appropriate for applications.
    
   Given that this section establishes an equivalence between the GJ set and BAOAB, it is perhaps intuitively clear that the numerical steady state distribution for each GJ method, which we denote as $\rho_{\rm GJ}$, should have weak second order convergence in the time step $\Delta t$, since this is known to be true in the case of BAOAB \cite{LM}. This intuitive picture is explicitly shown in the Appendix where, through the use of the Baker-Campbell-Hausdorff (BCH) formula, the weak second order nature of the limiting steady distribution, $\rho_{\rm GJ}$, is presented in a direct calculation. These  calculations agree with the result in Ref.~\cite{BIT}, which concluded that the GJF method is weakly second order, and extends this order property to \emph{all} GJ methods through the use of the splitting formalism we have developed. In addition to the formal argument given in the Appendix, we also numerically demonstrate this second order error scaling in the case of simple non-linear potentials for several GJ methods in the next section.

\section{Numerical Simulations}\label{sec:Num_Sim}

    \subsection{Simple Non-Linear System} \label{subsec:simple non-linear}
    Several previous publications have already demonstrated the statistical accuracy of the GJ methods for both configurational and kinetic measures for both simple nonlinear and complex molecular dynamics systems \cite{2GJ,GJ,Li2017,Josh,JDF2}. Of course, for linear systems these methods are designed to be statistically exact for any stable time step, but nonlinearities can create completely new dynamical features that dominate the system behavior for relatively large time steps \cite{2GJ2}. We here consider the more subtle nonlinearity induced, time step dependent deviations from the exact statistical expectations in continuous time by verifying the above-mentioned weak second order accuracy of the GJ set through simulations of simple, one-dimensional nonlinear oscillators. In order to assess the deviations from exact distributions we acquire moments $\langle (r^n)^k\rangle_n$ and $\langle(u^{n+\frac{1}{2}})^k\rangle_n$ of both configurational coordinate $r^n$ and its associated half-step velocity $u^{n+\frac{1}{2}}$ given by Eq.~(\ref{eq:GJ-u}). Given the equivalence between the methods that we have established in the previous section, we only show results for the GJ-I method (GJF-2GJ), as these are statistically indistinguishable from the comparable results obtained from any other GJ method. Two normalized potentials have been simulated:
    \begin{subequations}
    \begin{eqnarray}
    U_1(r) & = & \frac{1}{2}\kappa r^2(1+\kappa_{\rm nl}r^2)\label{eq:U1}\\
    U_2(r) & = & \frac{1}{2}\kappa r^2(1+\kappa_{\rm nl}(\frac{3}{2}r_0r+r^2))\label{eq:U2}\, ,
    \end{eqnarray}
    \end{subequations}
    where $\kappa\ge0$ is a spring constant, $\kappa_{\rm nl}$ adjusts the nonlinearity, and $r_0$ is the characteristic length-scale of the problem. Both quartic potentials have a single well with minima $U_i(0)=0$, the first well being symmetric, whereas the second is asymmetric and negatively skewed. The true, continuous-time values of the moments are given by the expected Boltzmann and Maxwell-Boltzmann distributions, respectively
    \begin{subequations}
    \begin{eqnarray}
        \left\langle (r)^k\right\rangle_{\rm exact} & = & \frac{\displaystyle\int_{-\infty}^\infty r^k\,e^{-U_i(r)/k_BT}\,dr}{\displaystyle\int_{-\infty}^{\infty}e^{-U_i(r)/k_BT}\,dr}\label{eq:moment_r_exact}\\
        \left\langle (u)^k\right\rangle_{\rm exact} & = & \frac{\displaystyle\int_{-\infty}^\infty u^k\,e^{-\frac{1}{2}mu^2/k_BT}\,du}{\displaystyle\int_{-\infty}^{\infty}e^{-\frac{1}{2}mu^2/k_BT}\,du} \; = \; \left(\frac{k_BT}{m}\right)^{\frac{k}{2}}\times\left\{\begin{array}{ccc}
        (k-1)!! & , & {\rm even } \; k>1 \\
        0 & , & {\rm odd} \; k\ge1
    \end{array}\right.\, ,
    \label{eq:moment_u_exact}
    \end{eqnarray}
    \end{subequations}
    where $\left\langle(r)^{2k-1}\right\rangle_{\rm exact}=0$ for the symmetric potential $U_1(r)$.

    Numerical simulations are conducted for the GJ-I method with the parameters
    \begin{subequations}
    \begin{eqnarray}
        k_BT & = & \kappa r_0^2\\
        \kappa_{\rm nl} r_0^2 & = & 1\\
        \gamma\Omega_0^{-1} & = & 1\, ,
    \end{eqnarray}\label{eq:sim_param}
    \end{subequations}
    where $\Omega_0=\sqrt{\kappa/m}$ is the natural frequency of the oscillator for $\kappa_{\rm nl}=0$, and time is normalized to $\Omega_0^{-1}$. For each reduced time step $\Delta{t}\Omega_0$, we then simulate $N=10^{11}$ time steps after an appropriate equilibration time, and we form the sample moments
    \begin{subequations}
    \begin{eqnarray}
        \left\langle(r^n)^k\right\rangle_n & = & \frac{1}{N}\sum_{n=1}^{N} (r^n)^k\label{eq:num_moment_r}\\
        \left\langle(u^{n+\frac{1}{2}})^k\right\rangle_n & = & \frac{1}{N}\sum_{n=1}^{N} (u^{n+\frac{1}{2}})^k\label{eq:num_moment_u}\, .
    \end{eqnarray}
    \end{subequations}
    We note that $\left\langle(r^n)^k\right\rangle_n\rightarrow\left\langle(r)^k\right\rangle_{\rm exact}$ and $\left\langle(u^{n+\frac{1}{2}})^k\right\rangle_n\rightarrow\left\langle(u)^k\right\rangle_{\rm exact}$ for $\kappa_{\rm nl}=0$ and $N\rightarrow\infty$ when using BAOAB or any of the GJ methods for $\kappa>0$.

    \begin{figure}[t]
    \centering
    \scalebox{0.7}{\centering \includegraphics[trim={1cm 2.0cm 1.0cm 8.0cm},clip]{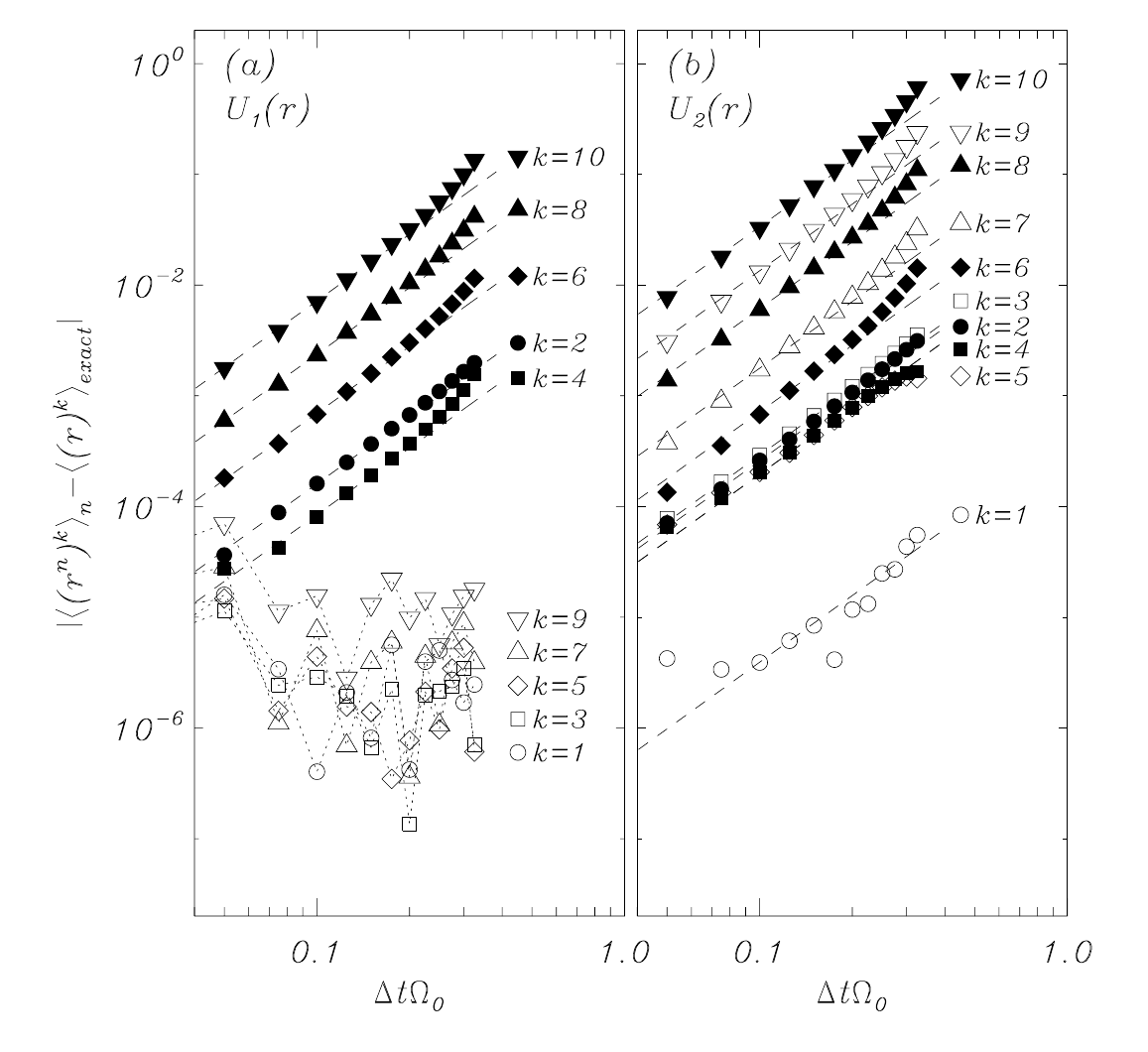}}
    \caption{Simulations of the error in the $k$th moment of the configurational coordinate $r^n$ as a function of reduced time step $\Delta{t}\Omega_0$ for a nonlinear oscillator with symmetric potential (a) given by Eq.~(\ref{eq:U1}), and asymmetric potential (b) given by Eq.~(\ref{eq:U2}). Each marker represents a GJ-I simulation, Eqs.~(\ref{eq:GJ}) and (\ref{eq:c2_gjf}), with $N=10^{11}$ time steps, and the resulting moment Eq.~(\ref{eq:num_moment_r}) is compared to the exact value given by Eq.~(\ref{eq:moment_r_exact}). Parameter values are given by Eq.~(\ref{eq:sim_param}). Dashed lines have slope 2. (a) Odd moments (open markers), expected to be zero for a symmetric potential, displayed with connecting dotted lines.}
    \label{fig:2nd_r}
    \end{figure}
    \begin{figure}[t]
    \centering
    \scalebox{0.7}{\centering \includegraphics[trim={1cm 2.0cm 1.0cm 8.0cm},clip]{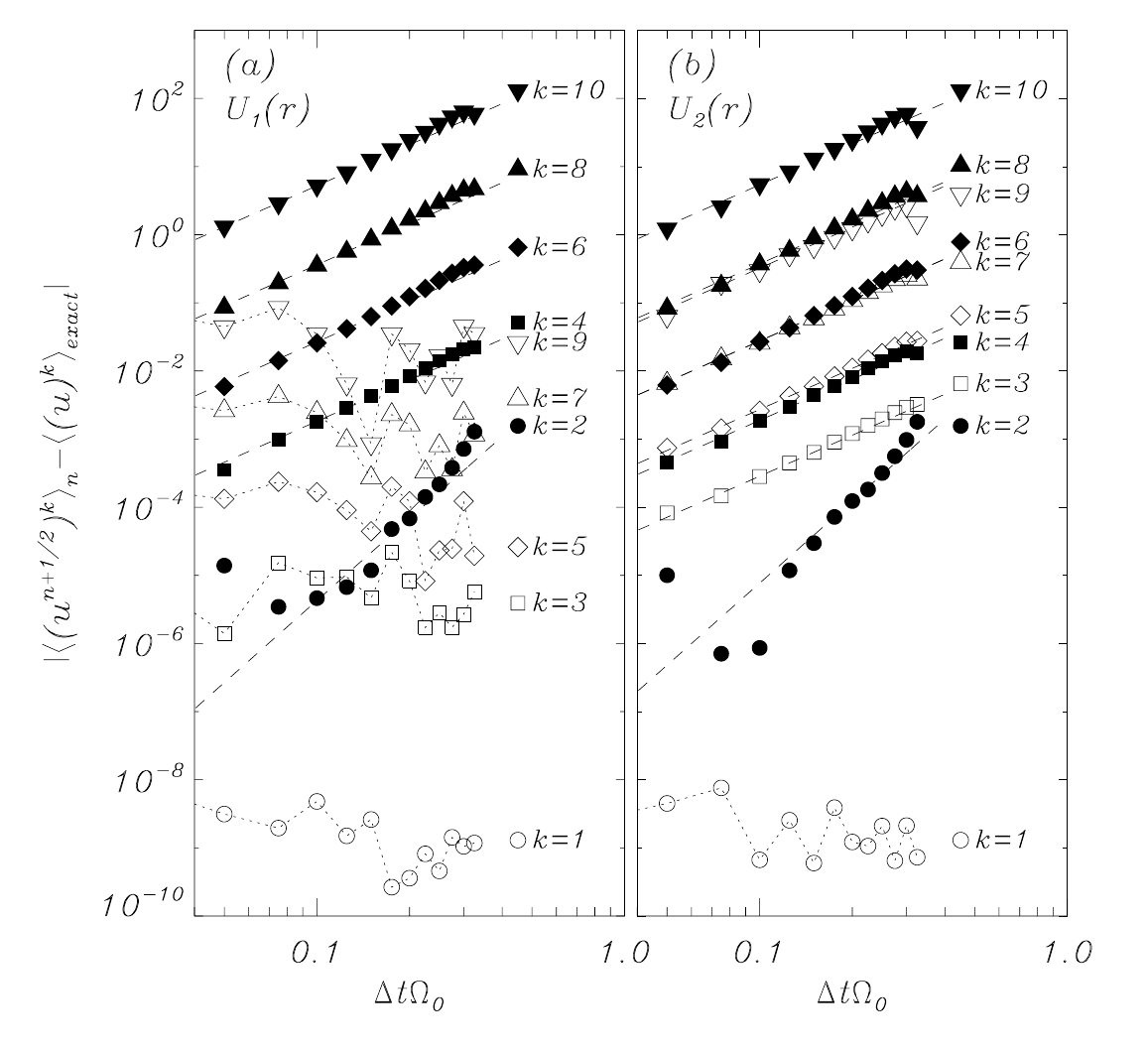}}
    \caption{Simulations of the error in the $k$th moment of the kinetic coordinate $u^{n+\frac{1}{2}}$ given by Eq.~(\ref{eq:GJ-u}) as a function of reduced time step $\Delta{t}\Omega_0$ for a nonlinear oscillator with symmetric potential (a) given by Eq.~(\ref{eq:U1}), and asymmetric potential (b) given by Eq.~(\ref{eq:U2}). Each marker represents a GJ-I simulation, Eqs.~(\ref{eq:GJ}) and (\ref{eq:c2_gjf}), with $N=10^{11}$ time steps, and the resulting moment Eq.~(\ref{eq:num_moment_u}) is compared to the exact value given by Eq.~(\ref{eq:moment_u_exact}). Parameter values are given by Eq.~(\ref{eq:sim_param}). Dashed lines have slope 2, except for the second moment, where the slope is 4. (a) Moments that are expected to be zero are displayed with connecting dotted lines.}
    \label{fig:2nd_u}
    \end{figure}

    Figures \ref{fig:2nd_r} and \ref{fig:2nd_u} show the differences between the non-zero continuous-time and discrete-time moments ($k\le10$) as a function of reduced time step $\Delta{t}\Omega_0$ for the GJ-I method applied to the two nonlinear oscillators with $\kappa_{\rm nl}=r_0^{-2}$. As described above, each marker signifies a simulation of $N=10^{11}$ time steps, and the simulated time steps span the entire stability range for the system. Figure \ref{fig:2nd_r} shows the results of the moments of the configurational degree of freedom along with dashed lines visualizing a quadratic slope. It is obvious that all the data are in very reasonable agreement with a second order error (in $\Delta{t}\Omega_0$) developing due to the nonlinearity $\kappa_{\rm nl}$ for both the symmetric (Fig.~\ref{fig:2nd_r}a) and the asymmetric (Fig.~\ref{fig:2nd_r}b) potentials, respectively.
  Ideally, by symmetry, all odd moments (open markers) in Fig.~\ref{fig:2nd_r}a should be zero regardless of $\Delta{t}$. However, lack of sampling efficiency prevents this, and the displayed open markers provide an indication of the statistical errors on calculating the moments. For example, we may infer the sampling error on the results for $k=6$ by looking at the magnitude of the displayed results for $k=5$ and $k=7$. We notice that the uncertainty (magnitude) of the odd moments tend to increase with the order of the moment. This is consistent with the fact that, for single well potentials, higher moments emphasize the statistical significance of infrequent events, thereby reducing the quality of the sampling for increasing $k$.
    Figure \ref{fig:2nd_u} similarly displays the comparable results for the kinetic, half-step degree of freedom. Again we observe very close agreement with second order error (in $\Delta{t}\Omega_0$) due to the nonlinearity $\kappa_{\rm nl}$ for both the symmetric (Fig.~\ref{fig:2nd_u}a) and the asymmetric (Fig.~\ref{fig:2nd_u}b) potentials, respectively. The noticeable discrepancy are the results for the $k=2$ order moments in Fig.~\ref{fig:2nd_u}, which behave with errors of seemingly fourth order in $\Delta{t}\Omega_0$. Of course, this is also in agreement with the expected weak (at least) second order discussed above.
  As for the configurational coordinate shown for the symmetric potential in Fig.~\ref{fig:2nd_r}a, Fig.~\ref{fig:2nd_u}a shows the odd moments (open markers) of the velocities being overall unaffected by the time step. By symmetry, these should be zero, but the sampling efficiency is poor for higher order moments. For the asymmetric potential in Fig.~\ref{fig:2nd_u}b, we also observe that the expected value of the first moment of the velocity is insignificant. Overall, Figs.~\ref{fig:2nd_r} and \ref{fig:2nd_u} show that for moments that are not guaranteed to be zero by symmetry, their time step dependent response to a nonlinearity is of (at least) second order in $\Delta{t}$.
    We re-emphasize that similar simulations for BAOAB and any other GJ method have, in agreement with the observation in Sec.~\ref{sec:equivalence}, produced statistically identical results to the ones shown here. The inconsistencies found in BAOAB do not appear for these studies since the simulated potentials prevent diffusion and transport.

\subsection{Quantum-based Molecular Dynamics} \label{sec:qmd}

    \begin{figure}[t]
    \centering
    \scalebox{0.7}{\centering \includegraphics[trim={0cm 2.0cm 1.0cm 8.0cm},clip]{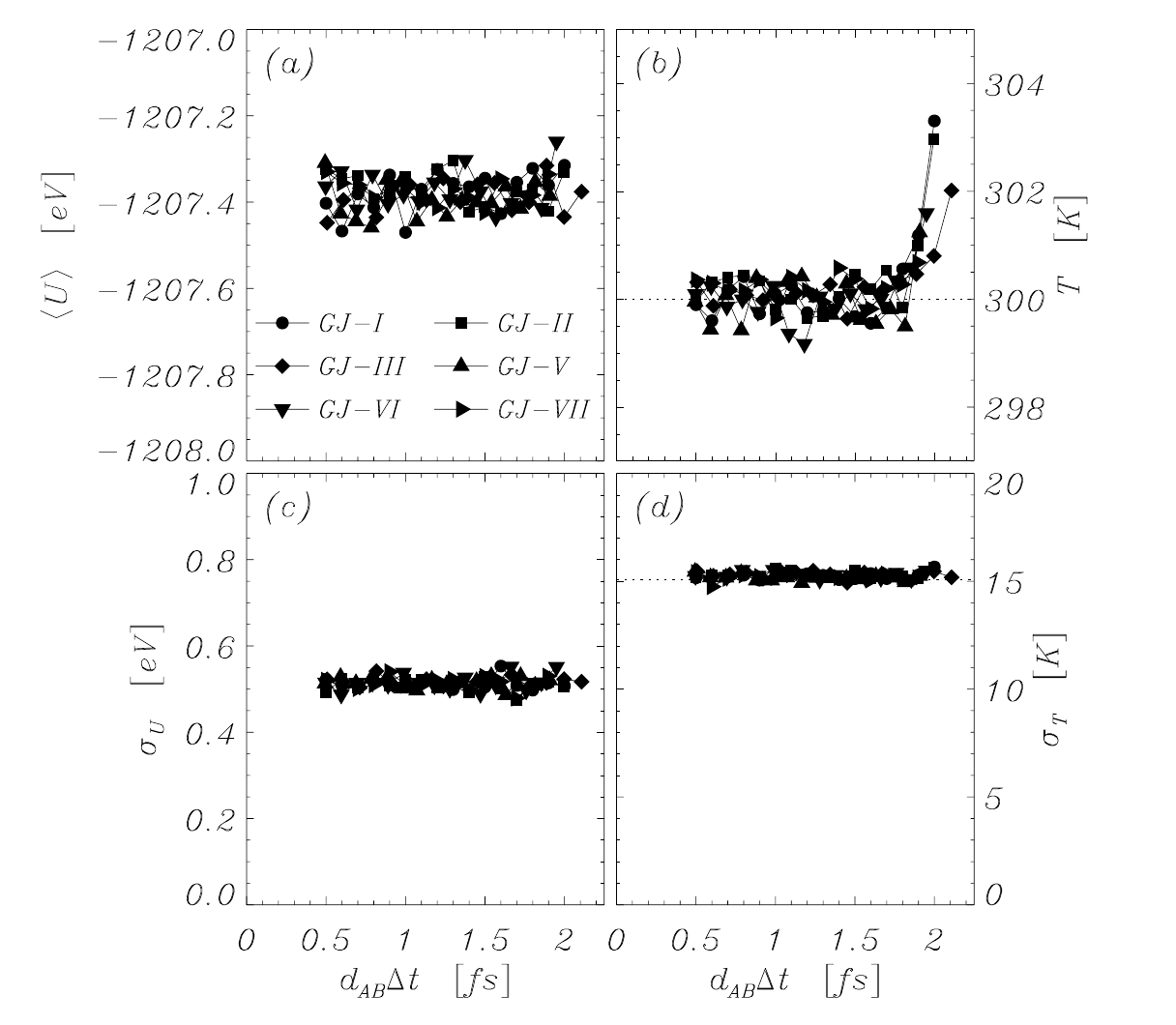}}
    \caption{Basic thermodynamic data as a function of time step from QMD LATTE \cite{LATTE} simulations at $T_0=300K$ with $88$ water molecules ($N=264$ atoms) using several GJ methods as indicated in panel (a). (a) Average potential energy of system $\langle U\rangle$, (b) kinetic temperature $T$ measured by the new half-step velocity given by Eq.~(\ref{eq:new_half_step}) (dashed line indicates $T_0$), (c) standard deviation $\sigma_U$ of the temporal fluctuations in the system potential energy, and (d) temporal fluctuations $\sigma_T$ in the instantaneous measure of system temperature, where the continuous-time value is given by $\sigma_T^2=2T_0/3N$ (dashed line). The stability limit of the time step is $d_{AB}\Delta{t}\lesssim2.0\,{\rm fs}$.}
    \label{fig:QMD}
    \end{figure}

    To further illustrate the utility of the GJ methods, we implemented several of these into the Los Alamos Transferable Tight-Binding for Energetics (LATTE) software package \cite{LATTE,PhysRevB.86.174308,perriot2018density}, an open source, self-consistent charge density-functional, tight-binding \cite{DFTB1,DFTB2,DFTB3,ANiklasson2015,Hourahine2020,Krishnapriyan2017} molecular dynamics code. The Fortran source code file used for the implementation is made available in the supplementary material. Calculations were performed in double-precision on Power9 compute nodes using an Nvidia Tesla V100 graphical processing unit (GPU). LATTE was compiled with MAGMA \cite{MAGMA} and cuSOLVER \cite{cusolver}, high-performance linear algebra software libraries specialized for GPUs, in order to take advantage of their fast linear algebra and diagonalization routines. 

    For this calculation we simulate a box of water with side lengths of 14 {\AA}  consisting of $N=264$ atoms (88 molecules). The simulation time was 20 ps and several different $\Delta t$ values were examined for the GJ--I, GJ--II, GJ--III, GJ--V, GJ--VI, and GJ--UTS (GJ-VII) methods. Since QMD simulations are far more expensive than classical MD simulations, we want to drive the system toward equilibrium as quickly as possible in order to obtain canonical sampling through the entire simulation run. This consideration often necessitates a large choice of friction coefficient and we therefore use the relatively large value of $\gamma = 10^{-1}$ fs$^{-1}$. This choice leads to the regime $\gamma \Delta t > 0.1$, which happens to also be where the methods most significantly differ. The new half-step velocity defined in \cref{eq:new_half_step} was calculated during the course of the simulations. At each time step, an electronic structure self-consistent field (SCF) iteration is performed to a pre-set tolerance of $10^(-6)$ in the difference of electron densities between iterations, a level of accuracy that ensures a negligible amount of drift in the total energy \cite{EMartinez2015}. Reaching this level of accuracy typically required about 5--10 iterations per time step, with each iteration requiring a dense matrix diagonalization of an electronic Hamiltonian \cite{ANiklasson2015} using cuSOLVER.

    The comparisons shown in Fig.~\ref{fig:QMD} clearly highlight the independence from the time step of the statistical errors in the potential and kinetic energy for QMD simulations with the relatively large choice of $\gamma = 10^{-1}$ fs$^{-1}$. Smaller friction values, though possible, would also generally require longer simulation time to reach an equilibrium distribution of states. This underlines the advantage of having a stochastic simulation method that produces robust statistical results for both large time steps \emph{and} large friction coefficients. We reemphasize that, within the sampling uncertainty, the depicted results show that statistical acquisition of both configurational and kinetic averages along with their fluctuations are constant over the near entirety of the stability range for the time step. The only noticeable departure from the expected result is a temperature increase by about 1\%  immediately prior to the instability ($1.8~\text{fs} < \Delta{} t < 2.1~\text{fs}$). Thus, we conclude from this set of simulations that the GJ methods can successfully describe thermodynamic properties in model systems with a quantum mechanical description of the interatomic forces.

\section{Discussion}
    This paper presents a direct connection between the most promising \cite{LM,Sivak} of the traditional Langevin operator splitting approaches and the brute force GJ approach \cite{GJ,JDF2}, which has led to the complete set of statistically sound stochastic Verlet-type Langevin integrators. It is concluded that the splitting approach alone cannot lead to a method that produces the same number of correct statistical properties as what is found in the GJ set. However, the ABO splitting of Ref.~\cite{LM} can be revised by a generalization of the time scaling that was proposed in Ref.~\cite{Sivak} for a particular choice of functional attenuation parameter $c_2$, and we are here able to determine the specific time scaling, $d_{AB}$ in Eq.~(\ref{eq:dAdB}), that brings the ABO splitting method into agreement with the entire GJ set.
    
    Reviewing the three discrete-time operator steps, A, B, and O, given in Eq.~(\ref{eq:Split_ABO}) (as well as in Eq.~(\ref{eq:Split_ABO_rev})), shows that the thermodynamic O operation (Eq.~(\ref{eq:Split_O})) is a correct representation of the fluctuation-dissipation balance, and that the ABO time step inconsistencies arise from the Euler operations in the inertial A (Eq.~(\ref{eq:Split_A})) and the interactive B (Eq.~(\ref{eq:Split_B})) steps. This is consistent with the observation that the necessary corrective measure applies to those two operations, while the thermodynamic O step can be revised by changing $c_2$ from one allowed functional form to another. Thus, the two approaches taken so far for correcting the transport properties of BAOAB can be illustrated by: 1) maintaining the time scaling of the A and B steps in BAOAB, while changing the attenuation parameter to the rational form $c_2=a$ in the thermodynamic O step, thereby obtaining GJ-I (GJF) \cite{GJF1}, and 2) maintaining the exponential form of the attenuation parameter $c_2$ in BAOAB's thermodynamic O step, while adjusting the time scaling $d_{AB}$ in BAOAB's inertial A and interactive B steps to match $c_2$, thereby obtaining GJ-II (VRORV) \cite{Sivak}. The entire GJ set can then arise from BAOAB by, e.g., choosing a time scaling $d_{AB}$ and then matching that scaling with an accompanying functional form for $c_2$ that produces correct statistics; or vice versa. However, the choices of the useful $c_2$ and $d_{AB}$ combination in the revision of BAOAB do not emerge naturally from the ABO splitting method, and must be determined by a direct parameter determination similar to the one conducted in the derivation of the GJ set of methods \cite{GJ,JDF2}.

    As illustrated in \cref{sec:general-splitting}, allowing departure from the exact exponential operators or inclusion of time scaling enables a structural gain --- namely correct (limiting) free particle and harmonic oscillator statistical behavior. This procedure can likely be generalized to other exact splittings in different orderings, which could possibly help to ``repair'' these methods by affording them the necessary flexibility to realize the desired statistical properties. This idea may prove fruitful whenever the algorithmic form of Eq.~(\ref{eq:gen_split}) is used, for example, in the numerical integration of quantum spin dynamics \cite{tranchida2018}. 

    Given the close connection between BAOAB and the GJ methods, we have been able to use the established analysis of accuracy for the LM family \cite{LM_book}, as shown in Appendix \ref{appx:appdx}, to the conclude that all the GJ methods are at least weakly second order accurate in the time step for nonlinear systems, consistent with the previously published analytical result for the GJF (GJ-I) method in Ref.~\cite{BIT}. The weak second order accuracy for the GJ methods also intuitively follows from the observation made in Section \ref{sec:equivalence}, that the GJ methods for any given time step can be described as one and the same method with different friction parameters.

    This paper takes further advantage of the ABO splitting analogy to develop a new and improved definition of the half-step velocity for {\it any} of the GJ methods. This velocity, shown in Eq.~(\ref{eq:new_half_step}), produces the correct kinetic statistics for the noisy harmonic oscillator {\it and} it produces the correct transport velocity. We have illustrated the integration of this new velocity into standard forms of the GJ methods. Although correct for the configurational coordinate in all GJ methods, the combination of correct velocity drift and Maxwell-Boltzmann statistics has previously only been possible for GJ-III \cite{GJ}, but ought to be available in all stochastic schemes. This is especially true whenever the Langevin equation is used for its original purpose of modeling the kinetic effects of an implicit solvent on solute atoms and molecules simulated explicitly. 
    
    In addition to kinetics, the unbiased estimation of atomic velocities is also key to the computation of certain \emph{thermodynamic} properties, such as the per-particle kinetic term of the virial stress \cite{Thompson2009}. There are several circumstances where a stochastic scheme with strong friction may be particularly beneficial. For example, when using many-body or quantum-mechanical interatomic potentials over short time scales \cite{Jensen2015}, or when simulating materials with a highly anisotropic distribution of forces and relaxation time scales \cite{Yeh2015,GFiorin21}.

    Apart from the now established mathematical connection between the ABO splitting and the GJ methods, the results of this paper have been numerically substantiated by simulating one-dimensional, nonlinear oscillators to verify the weak second order accuracy. 
        We also demonstrated the applicability of the GJ methods to condensed-phase systems by running molecular dynamics simulations at high friction for a small box of water described quantum mechanically. We show that the basic thermodynamic measures of potential and kinetic energies, along with their fluctuations, are robustly computed by the GJ methods within the entire stability range for the time step. These simulations have used the novel velocity definition presented in this paper, thus including the usability of this kinetic measure into the validation of the previously demonstrated statistically accurate GJ half-step velocities \cite{GJ}.

\section{Acknowledgements}
    This work was supported by the U.S. Department of Energy Office of Basic Energy Sciences (FWP LANLE8AN). Computing resources at Temple University were supported in part by the National Science Foundation through major research instrumentation grant \#1625061 and by the US Army Research Laboratory under contract \#W911NF-16-2-0189, the latter also supporting GF. BS was supported by National Science Foundation grants DMS--1952878 and DMS--2012271. We would also like to thank Christian~F.~A.~Negre for technical assistance, and the CCS-7 group and Darwin cluster at Los Alamos National Laboratory for computational resources. Darwin is funded by the Computational Systems and Software Environments (CSSE) subprogram of LANL's ASC program (NNSA/DOE). 

\section{Data Availability Statement}
    The data that support the findings of this study are available from the authors upon reasonable request.

\bibliography{bib}
\bibliographystyle{unsrt}

{\appendix

\section{Weak Second Order Accuracy}\label{appx:appdx}
\input{appendix}

}

\newpage
\begin{center}
{\bf{Supplementary Material}}
\end{center}
\input{SI}

\end{document}

%% file: appendix.tex
In this Appendix, we explicitly calculate the weak second order behavior for the GJ methods. The GJ splitting reformulation, as described in the main text, allows us to employ an established mathematical technique for operator splitting by using the operators defined in \cref{eq:Split_ABO_rev}. We take advantage of this formulation to formally certify the weak second order --- the order of convergence for averaged phase space quantities --- of each GJ method. As was done in Ref.~\cite{LM}, we use the Baker-Campbell-Hausdorff (BCH) formula to define a Fokker-Planck operator, $\L^*_{\text{GJ}}$, for each GJ method. We think of this operator as describing the evolution of the GJ methods' probability density in continuous time such that at each moment of discrete time, $t_n = n\Delta t$, the probability density of the numerical method is given by $e^{t_n \L^*_{\text{GJ}}} \rho_0$. Using the BCH formula, we define $\L^*_{\text{GJ}}$ through the following composition of linear operators
    \begin{align}
    \begin{split}\label{eq: define L_GJ star}
        \exp(\Delta t {{\L^*}_{\text{GJ}}}) & \equiv \exp(\tfrac{\Delta t}{2} \widetilde{\L^*}_{B})\exp(\tfrac{\Delta t}{2}{\widetilde{\L^*}}_{A})\exp({\Delta t}{\widetilde{\L^*}}_{O})
        \exp(\tfrac{\Delta t}{2}{\widetilde{\L^*}}_{A})\exp(\tfrac{\Delta t}{2}{\widetilde{\L^*}}_{B})\\
        & = \exp(\tfrac{d_{AB}\Delta t}{2}\L^*_B)\exp(\tfrac{d_{AB} \Delta t}{2}\L^*_A)\exp({d_{O} \Delta t}\L^*_O)
        \exp(\tfrac{d_{AB} \Delta t}{2}\L^*_A)\exp(\tfrac{d_{AB} \Delta t}{2}\L^*_B) \;,
    \end{split}
    \end{align}
    where each exponential operator on the right hand side is well-defined. Fixing $\gamma$ and expanding the time scaling factors in $\Delta t$, $d_{AB} = 1 + d_{AB}^{(1)}\Delta t + d_{AB}^{(2)} \Delta t^2 + \mathcal{O}(\Delta t^3)$ and $d_O = 1 + d_O^{(1)}\Delta t + d_O^{(2)} \Delta t^2 + \mathcal{O}(\Delta t^3)$, the $\L^*_{\text{GJ}}$ operator can be calculated up to second order in $\Delta t$,
    \begin{align}
    \begin{split}\label{eq: define L_GJ star - full}
        \L^*_{\text{GJ}} &= \L^*_{\text{LD}} + {\Delta t}(d_O^{(1)} \L^*_O + d_{AB}^{(1)}(\L^*_A+\L^*_B))\\
        &+ {\Delta t}^2(d_O^{(2)} \L^*_O + d_{AB}^{(2)}(\L^*_A+\L^*_B)) \\
        &+ \tfrac{\Delta t^2}{16} ( [\L^*_{\text{LD}},[\L^*_{O},\L^*_{A} + \L^*_B]] + [\L^*_{\text{LD}},[\L^*_{A},\L^*_{B}]] )  \\
        & + \tfrac{\Delta t^2}{48}([\L^*_A,[\L^*_A,\L^*_O]] + [\L^*_O,[\L^*_O,\L^*_A+\L^*_B]] + [\L^*_B,[\L^*_B,\L^*_A]] \\
        & + [\L^*_B,[\L^*_B,\L^*_O]] +  [\L^*_A,[\L^*_A,\L^*_B]]  \\
        & + [\L^*_O,[\L^*_A,\L^*_B]] + [\L^*_A,[\L^*_O,\L^*_B]]  )  +\mathcal{O}(\Delta t^3) \\
        & =: \L^*_{\text{LD}} + {\Delta t}\L^*_1 + {\Delta t}^2  \L^*_2
        + \L^*\;,
    \end{split}
    \end{align}
    where the result of applying the operator $\L^*$ scales as $\Delta t^3$ and $\L^*_{\rm LD} = \L^*_{\rm A}+\L^*_{\rm B}+\L^*_{\rm O}$ is the Langevin dynamics Fokker-Planck operator. For specific details on arriving to \cref{eq: define L_GJ star - full} from \cref{eq: define L_GJ star} please see the supplementary material. 
    Operating on the canonical distribution $\rho = \rho (r,v) \propto e^{-\beta \mathcal{H}(r,v)}$, leads to 
    \begin{align} \label{eq: L_star GJ}
        \L^*_{\text{GJ}} \; \rho &= \Delta t\L^*_1 \rho \; +\Delta t^2\L^*_2 \; \rho + \L^* \; \rho \; = \Delta t^2\L^*_2 \; \rho + \L^*\;\rho \;,
    \end{align}
    because $\L^*_1 \; \rho = 0$, since $\L^*_{\text{LD}} \; \rho = \L^*_O \; \rho = (\L^*_A+\L^*_B) \; \rho = 0$. The final equality follows from $\L^*_A+\L^*_B$ being the adjoint of the Liouvillian, which annihilates functions of the Hamiltonian. Therefore, up to second order in $\Delta t$, if $\L^*_2 \; \rho = \rho'$, we can re-write \cref{eq: L_star GJ} as 
    \begin{align} \label{eq: L_star GJ 2} 
        \L^*_{\text{GJ}} \; \rho = \Delta t^2 \; \rho' \;.
    \end{align}
    Then, assuming there exists a unique $\rho''$ such that we can solve the equation $\L^*_{\text{GJ}} \; \rho'' = \rho'$, we are able to then write \cref{eq: L_star GJ 2}  as
    \begin{align}
      \L^*_{\text{GJ}}(\underset{ \equiv \rho_{\rm GJ}}{\underbrace{\rho - \Delta t^2 \rho''}}) = 0 \;.
    \end{align}
    This shows that the steady-state distribution for the GJ methods, $\rho_{\rm GJ}$, satisfies $\rho_{\rm GJ} = \rho$ up to second order in $\Delta t$, i.e., $\rho_{\rm GJ} = \rho + \mathcal{O}(\Delta t^2)$, thereby demonstrating weak second order convergence in the time step if the potential is nonlinear. Of course, by construction there is no error in the distribution if the potential is linear. The above calculation further illuminates why it makes sense that the time scaling factors $d_A$ and $d_B$ of Eq.~(\ref{eq:dAdB}) are equal, since this allows for $d_{AB}^{(1)}(\L^*_A+\L^*_B)$ to annihilate $\rho$. A small calculation shows that $(d_{A}^{(1)}\L^*_A+d_{B}^{(1)}\L^*_B) \rho = 0$ implies $d_{A}^{(1)} = d_{B}^{(1)}$, so that a second-order method of the type in \cref{eq:gen_split} must have $d_{A}= d_{B}$ through first order in $\Delta t$.

%% file: SI.tex
\begin{center}
{\bf{I. LATTE Source Code for GJ Implementation}}
\end{center}

Here we include our modifications to the LATTE source file ``nvtlangevin.F90" that we used in order to implement the various GJ methods described in Sec.~\ref{sec:qmd}. The source files can be found on the github repository \url{https://github.com/lanl/LATTE}.\\

\lstset{language=[90]Fortran,
  basicstyle=\ttfamily,
  keywordstyle=\color{purple},
  backgroundcolor=\color{yellow!10},
  commentstyle=\color{cyan},
  morecomment=[l]{!\ }
}

\begin{lstlisting}
SUBROUTINE NVTLANGEVIN(ITER)

  USE CONSTANTS_MOD
  USE SETUPARRAY
  USE MDARRAY
  USE MYPRECISION

  IMPLICIT NONE

  INTEGER :: I, J, K, N
  INTEGER :: ITER
  REAL(LATTEPREC), EXTERNAL :: GAUSSRN
  REAL(LATTEPREC) :: MEAN, STDDEV, A, B, C, D, TEMPR(3,NATS)
  REAL(LATTEPREC) :: BOLTZ, SQMVV2KE
  REAL(LATTEPREC) :: GAMMA1, GAMMA2, TSQRT, DTF, DTV, DTFM
  REAL(LATTEPREC) :: PREF1, PREF2, MELPREF, MULTFACT, MYMASS
  INTEGER :: OPTION
  IF (EXISTERROR) RETURN

  SETTH = 0

  BOLTZ = 1.D0 / KE2T

  MEAN = ZERO

  GAMMA1 = 1.D0 / FRICTION

  !
  ! The functional parameter C2 for the GJ method.
  ! Given here is C=C2 for GJ-I (GJF).
  !
  C = (1.D0 - GAMMA1*DT/2.D0)/(1.D0 + GAMMA1*DT/2.D0) 

  !
  ! COMPUTE d_{AB}
  !
  D = SQRT(2.D0*(ONE-C)/(ONE+C)/GAMMA1/DT)  
  
  !
  ! HALF-STEP VELOCITY UPDATE
  !
  DO I = 1, NATS

     MULTFACT = (D*DT*F2V)/(TWO*MASS(ELEMPOINTER(I)))

     V(1,I) = V(1,I) + MULTFACT*FTOT(1,I) 
     V(2,I) = V(2,I) + MULTFACT*FTOT(2,I) 
     V(3,I) = V(3,I) + MULTFACT*FTOT(3,I) 

  ENDDO

  !
  ! STORE R_{N} FOR HALF-STEP VELOCITY UPDATE
  !
  TEMPR = CR

  !
  ! HALF-STEP POSITION UPDATE
  !
  CR = CR + D*DT*V/TWO

  !
  ! FULL-STEP ORNSTEIN-UHLENBECK UPDATE
  !
  DO I = 1, NATS

     STDDEV = SQRT(0.009648548D0*BOLTZ*300.D0*(1-C**2) \ 
              / MASS(ELEMPOINTER(I))) 
              ! CHANGE eV/amu TO (A/fs)^2
              ! 1 eV = 0.009648548D0 amu*(A/fs)^2

     V(1,I) = V(1,I)*C + GAUSSRN(MEAN,1.D0) * STDDEV
     V(2,I) = V(2,I)*C + GAUSSRN(MEAN,1.D0) * STDDEV
     V(3,I) = V(3,I)*C + GAUSSRN(MEAN,1.D0) * STDDEV

  ENDDO

  !
  ! HALF-STEP VELOCITY IMPLEMENTATION
  !
  A = (1.D0-C)/(GAMMA1*DT)
  HALFV = (CR-TEMPR)/DT 

  DO I = 1, NATS

     STDDEV = SQRT(2.D0*BOLTZ*300.D0*GAMMA1*DT) \        
              *SQRT((1-A)/2.D0/GAMMA1/DT/(MASS(ELEMPOINTER(I))) \ 
              /103.6509068D0) ! UNITS CONVERSION FROM amu TO A/fs 

     HALFV(1,I) = HALFV(1,I) + GAUSSRN(MEAN,1.D0) * STDDEV
     HALFV(2,I) = HALFV(2,I) + GAUSSRN(MEAN,1.D0) * STDDEV
     HALFV(3,I) = HALFV(3,I) + GAUSSRN(MEAN,1.D0) * STDDEV

  ENDDO

  !
  ! HALF-STEP POSITION UPDATE, GET R_{N+1}
  !
  CR = CR + D*DT*V/TWO

  !
  ! GET NEW FORCE, F_{N+1}
  !
  CALL GETMDF(1, ITER)

  !
  ! HALF-STEP VELOCITY UPDATE, GET V_{N+1}
  !
  DO I = 1, NATS

     MULTFACT = (D*DT*F2V)/(TWO*MASS(ELEMPOINTER(I)))

     V(1,I) = V(1,I) + MULTFACT*FTOT(1,I) 
     V(2,I) = V(2,I) + MULTFACT*FTOT(2,I) 
     V(3,I) = V(3,I) + MULTFACT*FTOT(3,I)

  ENDDO

  RETURN

END SUBROUTINE NVTLANGEVIN

\end{lstlisting}

\begin{center}
{\bf{II. Calculations in support of the Appendix}}
\end{center}

In this last section of the Supporting Information, we explicitly write out the calculations obtained from the definition of the $\L_\textrm{GJ}^*$ operator which were used in writing down Eq.~(\ref{eq: define L_GJ star - full}) in the appendix. Using the formula on pg. 287 of \cite{LM_book}, which is based on the BCH expansion, we define $\widetilde{\L^*_{OAB}}$ such that 
\begin{align}
    \begin{split}
     \exp \big(\tfrac{\Delta t}{2} \widetilde{\L^*_{OAB}}\big)
        & = \exp \big(\tfrac{\Delta t}{2} \widetilde{\L^*_{O}}\big) 
        \exp \big( \tfrac{\Delta t}{2} \widetilde{\L^*_{A}} \big) 
        \exp \big(\tfrac{\Delta t}{2} \widetilde{\L^*_{B}}\big)\\
        & = \exp\bigg( \tfrac{\Delta t}{2} \bigg( d_O\L^*_O + d_{AB}\L^*_A + d_{AB}\L^*_B\\
        &+ \tfrac{\Delta t}{4}(d_Od_{AB}[\L^*_O,\L^*_A] + d_Od_{AB}[\L^*_O,\L^*_B] + d_{AB}^2[\L^*_A,\L^*_B]) \\
        & +  \tfrac{\Delta t^2}{16}( d_Od_{AB}^2[\L^*_O,[\L^*_A,\L^*_B]])\\
        & + \tfrac{\Delta t^2}{48}(d_{AB}^3[\L^*_A,[\L^*_A,\L^*_B]] + d_{AB}^3[\L^*_B,[\L^*_B,\L^*_A]] \\
        &+ {d_O}^2d_{AB}[\L^*_O,[\L^*_O,\L^*_A]] + {d_O}^2d_{AB}[\L^*_O,[\L^*_O,\L^*_B]] \\
        & + d_O d_{AB}^2 [\L^*_A+\L^*_B,[\L^*_A+\L^*_B,\L^*_O]])  \bigg) \bigg) \\
    \end{split}
\end{align}
Similarly, we define $\widetilde{\L^*_{BAO}}$ such that 
\begin{align}
    \begin{split}
     \exp \big(\tfrac{\Delta t}{2} \widetilde{\L^*_{BAO}}\big)
        & = \exp \big(\tfrac{\Delta t}{2} \widetilde{\L^*_{B}}\big) 
        \exp \big( \tfrac{\Delta t}{2} \widetilde{\L^*_{A}} \big) 
        \exp \big(\tfrac{\Delta t}{2} \widetilde{\L^*_{O}}\big)\\
        & = \exp\bigg( \tfrac{\Delta t}{2} \bigg( d_{AB}\L^*_B + d_{AB}\L^*_A + d_O\L^*_O\\
        &+ \tfrac{\Delta t}{4}(d_{AB}^2[\L^*_B,\L^*_A] + d_Od_{AB}[\L^*_B,\L^*_O] + d_Od_{AB}[\L^*_A,\L^*_O]) \\
        & +  \tfrac{\Delta t^2}{16}( d_Od_{AB}^2[\L^*_B,[\L^*_A,\L^*_O]])\\
        &+ \tfrac{\Delta t^2}{48}(d_Od_{AB}^2[\L^*_A,[\L^*_A,\L^*_O]] + d_O^2d_{AB} [\L^*_O,[\L^*_O,\L^*_A]] \\
        &+ d_{AB}^3[\L^*_B,[\L^*_B,\L^*_A]] + {d_O}d_{AB}^2[\L^*_B,[\L^*_B,\L^*_O]] \\
        & + [d_{AB}\L^*_A+d_O\L^*_O,[d_{AB}\L^*_A+d_O\L^*_O,d_{AB}\L^*_B]])  \bigg) \bigg) \\
    \end{split}
\end{align}
Expanding the time scaling factors up to second order in $\Delta t$, with $d_O = 1 + d_O^{(1)}\Delta t + d_O^{(2)} \Delta t^2 + \mathcal{O}(\Delta t^3)$ and $d_{AB} = 1 + d_{AB}^{(1)}\Delta t + d_{AB}^{(2)} \Delta t^2 + \mathcal{O}(\Delta t^3)$, leads to
\begin{align}
\begin{split}
         \widetilde{\L^*_{OAB}} 
        &=  (1+d_O^{(1)}\Delta t + d_O^{(2)} \Delta t^2)\L^*_O + (1+d_{AB}^{(1)}\Delta t + d_{AB}^{(2)} \Delta t^2)(\L^*_A + \L^*_B)\\
        &+ \tfrac{\Delta t}{4}((1+d_{AB}^{(1)}\Delta t + d_{O}^{(1)}\Delta t)[\L^*_O,\L^*_A] + (1+d_{AB}^{(1)}\Delta t + d_{O}^{(1)}\Delta t)[\L^*_O,\L^*_B]\\
        &+ (1+2d_{AB}^{(1)}\Delta t)[\L^*_A,\L^*_B])\\
        &+  \tfrac{\Delta t^2}{16}([\L^*_O,[\L^*_A,\L^*_B]]) + \tfrac{\Delta t^2}{48}([\L^*_A,[\L^*_A,\L^*_B]] + [\L^*_B,[\L^*_B,\L^*_A]] \\
        &+ [\L^*_O,[\L^*_O,\L^*_A]] + [\L^*_O,[\L^*_O,\L^*_B]] + [\L^*_A+\L^*_B,[\L^*_A+\L^*_B,\L^*_O]]) \;.
        \end{split}
\end{align}
The zeroth order term is then
\begin{align}
    \L^*_{\text{LD}} = \L^*_O + \L^*_A + \L^*_B \;,
\end{align}
which is the Fokker-Plank operator for Langevin dynamics. The first order term is given by
\begin{align}
    \tfrac{1}{4}( 4 d_O^{(1)} \L^*_O + 4d_{AB}^{(1)}(\L^*_A+\L^*_B) +  [\L^*_O,\L^*_A] + [\L^*_O,\L^*_B] + [\L^*_A,\L^*_B]) \;,
\end{align}
and the $\Delta t^2$ term is given by
\begin{align}
    &\tfrac{1}{16}([\L^*_O,[\L^*_A,\L^*_B]] + 16d_O^{(2)} \L^*_O + 16d_{AB}^{(2)} (\L^*_A + \L^*_B)\\
    & + 4(d_{AB}^{(1)}+d_O^{(1)})[\L^*_O,\L^*_A+\L^*_B] + 8 d_{AB}^{(1)}[\L^*_A,\L^*_B]  )\\
    &+ \tfrac{1}{48}([\L^*_A,[\L^*_A,\L^*_B]] + [\L^*_B,[\L^*_B,\L^*_A]] \\
    &+ [\L^*_O,[\L^*_O,\L^*_A]] + [\L^*_O,[\L^*_O,\L^*_B]] + [\L^*_A+\L^*_B,[\L^*_A+\L^*_B,\L^*_O]]) \;.
\end{align}
Similarly, the reverse composition  $\widetilde{\L^*_{BAO}}$ leads to
\begin{align}
\begin{split}
         \widetilde{\L^*_{BAO}}  
        &=  (1+d_{AB}^{(1)}\Delta t + d_{AB}^{(2)} \Delta t^2)(\L^*_A + \L^*_B) + (1+d_O^{(1)}\Delta t + d_O^{(2)} \Delta t^2)\L^*_O \\
        &+ \tfrac{\Delta t}{4}((1+2d_{AB}^{(1)} \Delta t)[\L^*_B,\L^*_A] + (1 + d_{O}^{(1)}\Delta t + d_{AB}^{(1)}\Delta t )[\L^*_B,\L^*_O]\\
        &+ (1+d_{AB}^{(1)}\Delta t + d_{O}^{(1)}\Delta t)[\L^*_A,\L^*_O])\\
        &+  \tfrac{\Delta t^2}{16}([\L^*_B,[\L^*_A,\L^*_O]]  + \tfrac{\Delta t^2}{48}([\L^*_A,[\L^*_A,\L^*_O]] + [\L^*_O,[\L^*_O,\L^*_A]] \\
        &+ [\L^*_B,[\L^*_B,\L^*_A]] + [\L^*_B,[\L^*_B,\L^*_O]] + [\L^*_A+\L^*_O,[\L^*_A+\L^*_O,\L^*_B]]) \;,
        \end{split}
\end{align}
so that the low order terms are:
\begin{align}
    \Delta t^0 &:  \qquad \L^*_A + \L^*_B + \L^*_O \\
    \Delta t^1 &:  \qquad  \tfrac{1}{4} \big( 4d_{AB}^{(1)}(\L^*_A + \L^*_B) + 4d_O^{(1)} \L^*_O + [\L^*_B,\L^*_A] + [\L^*_B,\L^*_O] + [\L^*_A,\L^*_O] \big)                    \\
    \Delta t^2 &:  \qquad \tfrac{1}{16} \bigg( 16 d_{AB}^{(2)} (\L^*_A + \L^*_B) + 16 d_O^{(2)}\L^*_O + 8d_{AB}^{(1)} [\L^*_B,\L^*_A] \\
    & \qquad \qquad + 4(d_{AB}^{(1)} + d_O^{(1)})[\L^*_A + \L^*_B,\L^*_O] + [\L^*_B,[\L^*_A,\L^*_O]]  \bigg) \\
    &\qquad \qquad + \tfrac{1}{48} \bigg( [\L^*_A,[\L^*_A,\L^*_O]] + [\L^*_O,[\L^*_O,\L^*_A]] \\
    &\qquad \qquad + [\L^*_B,[\L^*_B,\L^*_A]] + [\L^*_B,[\L^*_B,\L^*_O]] + [\L^*_A+\L^*_O,[\L^*_A+\L^*_O,\L^*_B]])  \bigg)
\end{align}

With both $\widetilde{\L^*_{BAO}}$ and $\widetilde{\L^*_{OAB}}$ being written out, we can now calculate the GJ operator, $\L^*_{\text{GJ}}$, through second order in $\Delta t$ by once again applying the BCH formula,
    \begin{align}
    \begin{split}
        \exp(\Delta t \L^*_{\text{GJ}}) &= \exp(\tfrac{\Delta t}{2} \widetilde{\L^*_{BAO}}) \exp(\tfrac{\Delta t}{2} \widetilde{\L^*_{OAB}})\\
        & = \exp \bigg( \frac{\Delta t}{2} \bigg( \widetilde{\L^*_{BAO}}  + \widetilde{\L^*_{OAB}} + \frac{\Delta t}{4}[\widetilde{\L^*_{BAO}} , \widetilde{\L^*_{OAB}}] \\
        & + \frac{\Delta t^2}{48}([\widetilde{\L^*_{BAO}} ,[\widetilde{\L^*_{BAO}} , \widetilde{\L^*_{OAB}}]] - [ \widetilde{\L^*_{OAB}},[\widetilde{\L^*_{BAO}} , \widetilde{\L^*_{OAB}}]])  + \mathcal{O}(\Delta t^3) \bigg)\bigg) \\
        & = \exp \bigg( {\Delta t} \bigg( \frac{1}{2}(\widetilde{\L^*_{BAO}}  + \widetilde{\L^*_{OAB}}) + \frac{\Delta t}{8}[\widetilde{\L^*_{BAO}} , \widetilde{\L^*_{OAB}}] \\
        & + \frac{\Delta t^2}{96}([\widetilde{\L^*_{BAO}} ,[\widetilde{\L^*_{BAO}} , \widetilde{\L^*_{OAB}}]] - [ \widetilde{\L^*_{OAB}},[\widetilde{\L^*_{BAO}} , \widetilde{\L^*_{OAB}}]])  + \mathcal{O}(\Delta t^3) \bigg)\bigg) \;.
    \end{split}
    \end{align}
Calculating each term in the expansion on the right hand side,
\begin{align}
\begin{split}
    \widetilde{\L^*_{BAO}} + \widetilde{\L^*_{OAB}} &= 2 \L^*_{\text{LD}} + 2\Delta t(d_O^{(1)} \L^*_O + d_{AB}^{(1)}( \L^*_A + \L^*_B)\\
    &+ 2\Delta t^2 (d_O^{(2)} \L^*_O + d_{AB}^{(2)} (\L^*_A + \L^*_B) ) 
    + \tfrac{\Delta t^2}{16}([\L^*_O,[\L^*_A,\L^*_B]] + [\L^*_B,[\L^*_A,\L^*_O]])  \\
    &+ \tfrac{\Delta t^2}{48}([\L^*_A,[\L^*_A,\L^*_O]] + [\L^*_O,[\L^*_O,\L^*_A]] \\
    &+ [\L^*_B,[\L^*_B,\L^*_A]] + [\L^*_B,[\L^*_B,\L^*_O]] + [\L^*_A+\L^*_O,[\L^*_A+\L^*_O,\L^*_B]]  \\
    & +  [\L^*_A,[\L^*_A,\L^*_B]]  + [\L^*_B,[\L^*_B,\L^*_A]]  + [\L^*_O,[\L^*_O,\L^*_A]]\\
    & + [\L^*_O,[\L^*_O,\L^*_B]] + [\L^*_A+\L^*_B,[\L^*_A+\L^*_B,\L^*_O]]) +\mathcal{O}(\Delta t^3)\\ \\ 
    &= 2 \L^*_{\text{LD}} + 2\Delta t(d_O^{(1)} \L^*_O + d_{AB}^{(1)}( \L^*_A + \L^*_B)\\
    &+ 2\Delta t^2 (d_O^{(2)} \L^*_O + d_{AB}^{(2)} (\L^*_A + \L^*_B) ) 
    + \tfrac{\Delta t^2}{16}[\L^*_A,[\L^*_O,\L^*_B]]  \\
    &+ \tfrac{\Delta t^2}{48}(2[\L^*_A,[\L^*_A,\L^*_O]] + 2[\L^*_O,[\L^*_O,\L^*_A]] + 2[\L^*_B,[\L^*_B,\L^*_A]] \\
    & + 2[\L^*_B,[\L^*_B,\L^*_O]] +  2[\L^*_A,[\L^*_A,\L^*_B]]  + 2[\L^*_O,[\L^*_O,\L^*_B]] \\
    & + [\L^*_O,[\L^*_A,\L^*_B]] + [\L^*_B,[\L^*_A,\L^*_O]] ) +\mathcal{O}(\Delta t^3) \\ \\
    &= 2 \L^*_{\text{LD}} + 2\Delta t(d_O^{(1)} \L^*_O + d_{AB}^{(1)}( \L^*_A + \L^*_B)\\
    &+ 2\Delta t^2 (d_O^{(2)} \L^*_O + d_{AB}^{(2)} (\L^*_A + \L^*_B) ) \\
    &+ \tfrac{\Delta t^2}{48}(2[\L^*_A,[\L^*_A,\L^*_O]] + 2[\L^*_O,[\L^*_O,\L^*_A]] + 2[\L^*_B,[\L^*_B,\L^*_A]] \\
    & + 2[\L^*_B,[\L^*_B,\L^*_O]] +  2[\L^*_A,[\L^*_A,\L^*_B]]  + 2[\L^*_O,[\L^*_O,\L^*_B]] \\
    & + 2[\L^*_O,[\L^*_A,\L^*_B]] + 2[\L^*_A,[\L^*_O,\L^*_B]]  ) +\mathcal{O}(\Delta t^3)
\end{split}
\end{align}
and
\begin{align}
\begin{split}
    [\widetilde{\L^*_{\text{BAO}}},\widetilde{\L^*_{\text{OAB}}}] &= [\L^*_{\text{LD}},\L^*_{\text{LD}}] + \Delta t(d_O^{(1)}[\L^*_{\text{LD}},\L^*_{O}] + d_{AB}^{(1)}[\L^*_{\text{LD}},\L^*_{A} + \L^*_{B}]\\
    &+ \tfrac{1}{4}([\L^*_{\text{LD}},[\L^*_{O},\L^*_{A}]] + [\L^*_{\text{LD}},[\L^*_{O},\L^*_{B}]]+[\L^*_{\text{LD}},[\L^*_{A},\L^*_{B}]]))\\
    & + \Delta t(d_O^{(1)}[\L^*_{O},\L^*_{\text{LD}}] + d_{AB}^{(1)}[\L^*_{A}+\L^*_{B},\L^*_{\text{LD}}]  \\
    &+ \tfrac{1}{4}([[\L^*_{A},\L^*_{O}],\L^*_{\text{LD}}] + [[\L^*_{B},\L^*_{O}],\L^*_{\text{LD}}]+[[\L^*_{B},\L^*_{A}],\L^*_{\text{LD}}])) +\mathcal{O}(\Delta t^2)\\ \\
    &= \tfrac{\Delta t}{2}( [\L^*_{\text{LD}},[\L^*_{O},\L^*_{A} + \L^*_B]] + [\L^*_{\text{LD}},[\L^*_{A},\L^*_{B}]] )
    +\mathcal{O}(\Delta t^2) \;,
\end{split}
\end{align}
and, 
\begin{align}  
    [\L^*_{{BAO}},[\L^*_{{BAO}},\L^*_{{OAB}}]] - [\L^*_{{OAB}},[\L^*_{{BAO}},\L^*_{{OAB}}]] =  \mathcal{O}(\Delta t)\;,
\end{align}
since only zero order terms are iterated commutators of $\L^*_{\rm LD}$. Therefore, up to second order in $\Delta t$,
\begin{align}
\begin{split} \label{L_GJ}
    \L^*_{\text{GJ}} &= \L^*_{\text{LD}} + {\Delta t}(d_O^{(1)} \L^*_O + d_{AB}^{(1)}(\L^*_A+\L^*_B))\\
    &+ {\Delta t}^2(d_O^{(2)} \L^*_O + d_{AB}^{(2)}(\L^*_A+\L^*_B)) \\
    &+ \tfrac{\Delta t^2}{16} ( [\L^*_{\text{LD}},[\L^*_{O},\L^*_{A} + \L^*_B]] + [\L^*_{\text{LD}},[\L^*_{A},\L^*_{B}]] )  \\
    & + \tfrac{\Delta t^2}{48}([\L^*_A,[\L^*_A,\L^*_O]] + [\L^*_O,[\L^*_O,\L^*_A]] + [\L^*_B,[\L^*_B,\L^*_A]] \\
    & + [\L^*_B,[\L^*_B,\L^*_O]] +  [\L^*_A,[\L^*_A,\L^*_B]]  + [\L^*_O,[\L^*_O,\L^*_B]] \\
    & + [\L^*_O,[\L^*_A,\L^*_B]] + [\L^*_A,[\L^*_O,\L^*_B]]  )  +\mathcal{O}(\Delta t^3)\;.
\end{split}
\end{align}